\documentclass[11pt]{article}
\usepackage[a4paper,margin=2.5cm]{geometry}
\usepackage{amsmath,amssymb,amsthm}
\usepackage{thmtools}
\usepackage[round,authoryear]{natbib}
\usepackage{booktabs,array}
\usepackage{graphicx}
\usepackage{xurl}
\usepackage{xr-hyper}
\usepackage[hidelinks,hyperfootnotes=false]{hyperref}
\usepackage{setspace}
\usepackage[section]{placeins}
\newtheorem{lemma}{Lemma}
\newtheorem{corollary}{Corollary}
\newtheorem{alemma}[lemma]{Lemma}
\newtheorem*{lemmaone}{Lemma~\ref{lem:rep}}
\theoremstyle{definition}
\newtheorem{definition}{Definition}

\theoremstyle{remark}
\newtheorem{remark}{Remark}

\makeatletter
\renewcommand{\@fnsymbol}[1]{\ensuremath{\ifcase#1\or\ast\or\dagger\or\ddagger\else\@ctrerr\fi}}
\makeatother

\newcommand{\E}{\mathbb{E}}

\newcommand{\dd}{\,\mathrm{d}}
\newcommand{\yb}{\bar y}
\newcommand{\gb}{\bar g}

\title{Interpreting the Morbidity Gap}
\author{Alec Morton\thanks{Address for correspondence: Alec Morton, Saw Swee Hock School of Public Health, National University of Singapore, 12 Science Drive 2, Singapore 117549, Singapore. E-mail: \href{mailto:a.morton@nus.edu.sg}{a.morton@nus.edu.sg}}\\[4pt] \small Saw Swee Hock School of Public Health, National University of Singapore}
\date{\today}

\begin{document}
\maketitle

\renewcommand{\abstractname}{Summary}
\begin{abstract}
\noindent The Global Burden of Disease (GBD) study now reports the morbidity gap (MG), life expectancy minus healthy life expectancy, as an indicator in its own right. Writing MG as the expected cumulative rate of years lived with disability (YLDs) at the life-table age at death, we use stochastic dominance results to show what MG measures, how it responds to mortality and how it relates to the YLDs of the current population. We show the following. (a) A YLD profile gives an MG no larger than another's under every life table if and only if its cumulative YLD rate is nowhere higher. (b) With YLD rates fixed, lower mortality at every age widens MG, so that removing a risk factor that raises mortality reduces MG by less than its contribution to MG in GBD's risk factor attribution. (c) A family of results gives easily interpretable conditions on the age structure of the population for the crude YLD rate to be above or below the life-table YLD rate, which is MG divided by life expectancy, and these conditions change with population ageing. We illustrate the results for Singapore, where MG widened from 8.9 to 11.2 years between 1990 and 2023.

\end{abstract}

\newpage
\section{Introduction}

The Global Burden of Disease (GBD) now treats the difference between life expectancy and healthy life expectancy (HALE) -- the so-called ``Morbidity Gap'' (MG) -- as an indicator in its own right and attributes it to diseases, injuries and risk factors  \citep{hay2026}.  \citet{garmany2024,garmany2025} have compared the same quantity across the member states of the World Health Organization, calling it the healthspan--lifespan gap. A widening MG is naturally read as evidence that non-fatal health is getting worse. In Singapore, for example, MG has widened from 8.9 years in 1990 to 11.2 years in 2023, with high fasting plasma glucose and high body-mass index the leading contributing risk factors \citep{hay2026,nus2026,moh2026}. Since GBD computes HALE by Sullivan's method, MG is the sum over ages of life-table person-years per birth times the age-specific rate of years lived with disability (YLDs) \citep{salomon2012}. The same rates, applied to the people alive in a year, give the YLDs that GBD reports for that year, which we call cross-sectional YLDs (CY).

In this paper, we ask what MG measures, how it compares across mortality schedules when age-specific YLD rates are held fixed,  and how it is related to CY. Since, as we show, MG is the expected cumulative YLD rate at the life-table age at death (Lemma~\ref{lem:rep}), comparisons of MG have the structure of comparisons of expected utility, and classical stochastic dominance results address them \citep{hadar1969,rothschild1970,shaked2007}. Both MG and CY average a curve of YLD rates over a distribution of age, whether a life table or a population. Dominance conditions show how relations between particular summary numerical indicators such as MG and CY are often reflective of the deeper and more fundamental  structural relationships of the distributions themselves.

Our results specialise standard characterisations of stochastic orders.  The novelty is in the light that they shed on the properties of MG as an indicator. Section~\ref{sec:lit} reviews the literature on which we draw. Section~\ref{sec:gap} deals with comparisons of MG between distributions which differ only in respect of morbidity and between distributions which differ only in respect of mortality, and Section~\ref{sec:cross} with comparisons between MG and CY. Section~\ref{sec:illustration} applies the results to Singapore.  Section~\ref{sec:conclusion} concludes.   The proofs are in Appendix~\ref{app:proofs}, and the construction of the figures is in Appendix~\ref{app:figs}.

\section{Related literature}\label{sec:lit}

MG is a relatively new indicator.  However, the debate on whether falling mortality compresses or expands morbidity has been going on for some time. The view of \citet{gruenberg1977} was that treatments that prevent the fatal complications of chronic disease, such as the sulfonamides that cut deaths from pneumonia, prolong the lives of people who have such diseases and so expand morbidity.  \citet{kramer1980} added a second mechanism, the growth of the age groups at high risk, which raises crude prevalence even when age-specific prevalence is unchanged. On the other hand, the hypothesis of a compression of morbidity is due to \citet{fries1980}, who argued that, with the length of life approaching a biological limit, postponing the onset of chronic illness would compress morbidity into a shorter period before a nearly fixed age at death. \citet{manton1982} proposed a dynamic equilibrium in which falling mortality is accompanied by slower progression of chronic disease, so that there is more disease but with milder consequences \citep{crimmins2011}, and he warned that age-specific health could improve while the absolute demand for services rose with the size of the elderly population.

Health expectancies, which are usually computed by Sullivan's method, have become a favoured analysis tool in this debate. Studies that delete causes from life tables show how mortality and morbidity interact: eliminating a highly fatal disease such as cancer adds years lived with disability, because the people it would have killed survive to ages at which disability from other causes is common, whereas eliminating a disabling but rarely fatal condition such as arthritis removes years lived with disability \citep{nusselder1996,mathers1999}. The evidence on trends mostly points to expansion: in the United States the length of life with major disease and with loss of mobility increased between 1998 and 2008 \citep{crimmins2011}; in GBD's estimates each additional year of life expectancy at birth came with about ten months of HALE between 1990 and 2010 \citep{salomon2012}; MG widened between 1990 and 2023 in all but one of 204 countries and territories and is larger where life expectancy is higher \citep{hay2026}.   \citet{garmany2024,garmany2025} find in the World Health Organization's estimates for 183 countries that it widened between 2000 and 2019 and that life expectancy is among its strongest correlates. The studies of trends read a widening MG as evidence about non-fatal health, and they leave open how much of the widening the fall in mortality would have produced on its own, with age-specific YLD rates unchanged.

Health economists have generally been critical of GBD's measures.  Their objections, to the age weights and discounting of the original DALY \citep{anand1997}, which GBD dropped in 2010 \citep{murray2012}, to the meaning of the disability weights \citep{nord2013} and to the use of totals of burden in setting priorities \citep{williams1999,mooney2000}, concern questions that this paper does not address \citep[for a defence of the DALY, see][]{murray1997}. An exception is \citet{airoldi2009}, who showed that when years of life lost are counted against life expectancy at the age of death, extending a life can increase the measured burden of disease, so that a planner who minimises disease burden can refuse a treatment that a planner who maximises health would provide. MG has a similar property, since with age-specific YLD rates fixed every year of life that falling mortality adds is lived with some health loss and so widens MG. Recently \citet{morenoternero2023} have brought the debate to a higher level by developing an axiomatic framework which shows how both DALYs and the more familiar QALYs are members of the same family of lifecourse health measures.

Dominance methods, which health economists use routinely in measuring inequality in health, have not been brought to bear on GBD's measures. Dominance conditions are used to check whether a ranking holds for every index in a family \citep{odonnell2008,makdissi2014}. The same logic underlies the dominance results of inequality and poverty measurement \citep{atkinson1970,foster1988} and their applications to ordinal health data \citep{allison2004}, to several dimensions of population health \citep{hussain2016} and to distributions of age at death \citep{issa2024}. GBD's measures, by contrast, are reported as single numbers computed from one life table and one standard population, although the ranking of causes by age-standardised rates can change with the standard \citep{wyper2020}.

In this paper we show how stochastic dominance results can inform the interpretation of MG. Each of our results is a ranking of the robust kind, holding for every life table or for every rising YLD profile, and its condition compares cumulative curves, as in the concentration index of inequality measurement  \citep{wagstaff1991,kakwani1997,odonnell2008}, with age in place of income rank.

\section{The meaning and behaviour of MG}\label{sec:gap}

\subsection{Definitions}\label{sec:notation}

Lower-case letters denote rates and densities by age and upper-case letters their cumulative versions, so that $Y$ and $C$ accumulate $y$ and $c$; $S$ and $T$ keep their life-table names. A bar denotes an average over a population. Superscripts such as $A$ and $B$ label populations, with $G^A=G(S^A,y^A)$. The operator $\E$ is used only for expectations over the life-table age at death $X$. Table~\ref{tab:notation} lists the symbols and where each is defined. We start, as GBD does, from the age structure, mortality rates and YLD (Years Lived with Disability) rates of a population and build its life table from the mortality rates; the definitions of the life table and the life-table population are standard (\citealp[ch.~3]{preston2001}; \citealp[ch.~2]{keyfitz2005}).

\begin{definition}[Population]\label{def:pop}
A \emph{population} at a given time, with ages in $[0,\omega]$ for some $\omega<\infty$, has an age density $c$ with distribution function $C$, so that $C(a)$ is the share of the population younger than $a$; and a mortality rate $\mu(a)$ at each age $a$.  The population also has a \emph{YLD profile} $y$, defined as a function on $[0,\omega]$ with values in $[0,1]$. $y(a)$ represents as all-cause YLDs per person-year at age $a$ and can be interpreted as the mean health loss of people alive at that age. The \emph{cumulative YLD rate} is
\begin{equation}\label{eq:Y}
Y(a)=\int_0^ay .
\end{equation}
\end{definition}

\begin{definition}[Life table]\label{def:lifetable}
The \emph{life table} of a population is
\begin{equation}\label{eq:S}
S(a)=\exp\Big\{-\int_0^a\mu(v)\dd v\Big\},
\end{equation}
the probability of surviving to age $a$ at the population's current mortality rates. More generally, a \emph{life table} is any function $S$ on $[0,\omega]$ that is nonincreasing and right-continuous with $S(0)=1$ and $S(\omega)=0$; results that hold for every life table refer to this wider class, which includes life tables without a mortality rate, such as one in which everyone dies at the same age. The life-table \emph{age at death} is the random variable $X$ with
\begin{equation}\label{eq:X}
\Pr(X>a)=S(a)\quad\text{for every $a\in[0,\omega]$.}
\end{equation}
Life expectancy at birth is
\begin{equation}\label{eq:e0}
e_0=\E[X]=\int_0^\omega S(v)\dd v,
\end{equation}
and person-years lived above age $a$ are
\begin{equation}\label{eq:T}
T(a)=\int_a^\omega S(v)\dd v .
\end{equation}
The \emph{life-table population} has age density $S/e_0$ and, by \eqref{eq:T}, distribution function
\begin{equation}\label{eq:ltpop}
\int_0^a\frac{S(v)}{e_0}\dd v=1-\frac{T(a)}{e_0},
\end{equation}
so that the share of its members older than age $a$ is $T(a)/e_0$.
\end{definition}

\begin{definition}[MG, life-table and crude YLD rates, and CY]\label{def:mg}
For a life table $S$ and YLD profile $y$, Sullivan's method, which GBD applies to life-table person-years and comorbidity-adjusted YLDs per person \citep[appendix~1, pp.~62--63]{gbd2023dirf}, gives HALE as $\int S(1-y)$.  MG, life expectancy minus HALE, is
\begin{equation}\label{eq:G}
G(S,y)=\int_0^\omega S(a)y(a)\dd a
\end{equation}
years per birth, written $G$ when $S$ and $y$ are clear. For an age $a$ with $S(a)>0$, \emph{MG from age $a$} is the corresponding quantity for the people alive at age $a$,
\begin{equation}\label{eq:Ga}
G_a(S,y)=\frac{1}{S(a)}\int_a^\omega S(v)y(v)\dd v .
\end{equation}
The \emph{life-table YLD rate} (LYR) is MG per year of life expectancy,
\begin{equation}\label{eq:gbar}
\gb(S,y)=\frac{G}{e_0}=\int_0^\omega \frac{S(a)}{e_0}\,y(a)\dd a=1-\frac{\mathrm{HALE}}{e_0},
\end{equation}
which is the average YLD rate over the life-table population. The \emph{crude YLD rate} (CYR) is the average YLD rate over the population itself,
\begin{equation}\label{eq:yb}
\yb=\int_0^\omega c(a)y(a)\dd a ,
\end{equation}
which GBD reports as the all-ages YLD rate. The YLDs of the people alive in a year, which GBD reports as the number of YLDs, are the size of the population times $\yb$, and we call them the \emph{cross-sectional YLDs} (CY) of the population.
\end{definition}

MG and the LYR depend on a population only through its mortality rates and its YLD rates, whereas the CYR and CY also depend on its age structure.

\begin{table}[!htbp]
\centering
\small
\caption{Notation. Symbols used only within a single result or proof are defined where they appear.}\label{tab:notation}
\begin{tabular}{@{}>{\raggedright}p{2.6cm}>{\raggedright}p{9.0cm}>{\raggedright}p{2.9cm}@{}}
\toprule
Symbol & Meaning & Defined in\tabularnewline
\midrule
$c(a)$ & Age density of the population & Def.~\ref{def:pop}\tabularnewline
$C(a)$ & Distribution function of the population: the share younger than $a$ & Def.~\ref{def:pop}\tabularnewline
$\mu(a)$ & Mortality rate at age $a$ & Def.~\ref{def:pop}\tabularnewline
$y(a)$ & YLD rate at age $a$ (the YLD profile) & Def.~\ref{def:pop}\tabularnewline
$Y(a)$ & Cumulative YLD rate & \eqref{eq:Y}\tabularnewline
$S(a)$ & Life table: probability of surviving to age $a$ & Def.~\ref{def:lifetable}, \eqref{eq:S}\tabularnewline
$X$ & Age at death in the life table & \eqref{eq:X}\tabularnewline
$e_0$ & Life expectancy at birth & \eqref{eq:e0}\tabularnewline
$T(a)$ & Person-years lived above age $a$; $T(a)/e_0$ is the share of the life-table population older than $a$ & \eqref{eq:T}, \eqref{eq:ltpop}\tabularnewline
$G$ & MG: life expectancy minus HALE, in years & \eqref{eq:G}\tabularnewline
$G_a$ & MG from age $a$ & \eqref{eq:Ga}\tabularnewline
$\gb$ & LYR: the life-table YLD rate, MG per year of life expectancy, which is the average YLD rate over the life-table population & \eqref{eq:gbar}\tabularnewline
$\yb$ & CYR: the crude YLD rate, cross-sectional YLDs (CY), the YLDs of the people alive in a year, per head & \eqref{eq:yb}\tabularnewline
$\Gamma_{\mathrm{mb}}$ & Morbidity component of a risk factor's deletion effect, which is GBD's contribution of the risk factor to MG & \eqref{eq:Gamma}\tabularnewline
$\Gamma_{\mathrm{mt}}$ & Mortality component of a risk factor's deletion effect & \eqref{eq:Gamma}\tabularnewline
OTILT & Older than its life-table population: $1-C\ge T/e_0$ at every age & Def.~\ref{def:otilt}\tabularnewline
YTILT & Younger than its life-table population: $1-C\le T/e_0$ at every age & Def.~\ref{def:otilt}\tabularnewline
MOYTILT & More old years than its life table: at least as many years per head lived beyond every age as the life-table population & \eqref{eq:moytilt}\tabularnewline
FOYTILT & Fewer old years than its life table: at most as many years per head lived beyond every age as the life-table population & \eqref{eq:foytilt}\tabularnewline
\bottomrule
\end{tabular}
\end{table}

\begin{remark}[Abridged tables]\label{rem:abridged}
In GBD's abridged tables MG is $\sum_x{}_nL_x\,y_x$; all results hold with sums in place of integrals, and conditions on piecewise-linear functions such as $Y$ need checking only at age-group boundaries.
\end{remark}

\begin{lemma}[MG as an expectation over age at death]\label{lem:rep}
For every life table $S$ and YLD profile $y$,
\begin{equation}\label{eq:rep}
G=\E[Y(X)]
\end{equation}
and, for every $a$ with $S(a)>0$,
\begin{equation}\label{eq:repa}
G_a=\E[Y(X)-Y(a)\mid X>a].
\end{equation}
\end{lemma}

Since $Y$ is nondecreasing, and convex when $y$ is nondecreasing, comparing MG across life tables for a given YLD profile compares expected values of a nondecreasing (and possibly convex) function of age at death, the setting of the classical dominance theorems.

\subsection{Comparing MG when only YLD rates differ}\label{sec:common-mort}

Comparing MG across YLD profiles with mortality held fixed needs no assumption about how morbidity and mortality interact, since the two populations being compared share a life table and differ only in their age-specific YLD rates. The question is when one YLD profile gives a smaller MG than another whatever the life table, and Theorem~\ref{thm:common-mort} answers this question with a condition on the cumulative YLD rates alone.

\begin{restatable}[Common mortality]{theorem}{thmcommonmort}\label{thm:common-mort}
For two YLD profiles $y^A$ and $y^B$, the following are equivalent:
\begin{enumerate}
\item[(i)] $G(S,y^A)\le G(S,y^B)$ for every life table $S$;
\item[(ii)] $\gb(S,y^A)\le\gb(S,y^B)$ for every life table $S$;
\item[(iii)] $Y^A(a)\le Y^B(a)$ for every age $a$.
\end{enumerate}
Moreover, $G_a(S,y^A)\le G_a(S,y^B)$ for every life table $S$ and every $a$ with $S(a)>0$ if and only if $y^A(a)\le y^B(a)$ at every age.
\end{restatable}

\begin{figure}[!htbp]
\centering
\includegraphics[width=\textwidth]{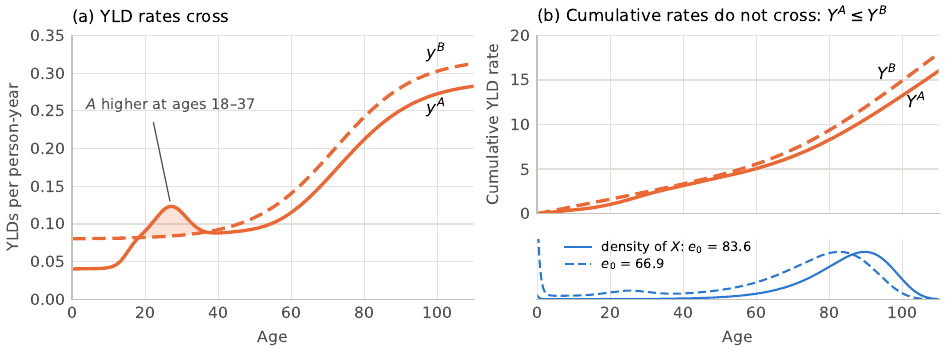}
\caption{Two synthetic YLD profiles. (a) YLD profiles $y^A$ and $y^B$. (b) Cumulative YLD rates $Y^A$ and $Y^B$, with the densities of age at death $X$ under two life tables beneath.}\label{fig:thm-common}
\end{figure}

Since the two populations share a life table, and so a life expectancy, conditions (i) and (ii) are equivalent; (ii) is stated to bring out the parallel with Theorem~\ref{thm:mortality}, in which the conditions on MG and on the LYR differ. Condition (iii) is weaker than $y^A\le y^B$, so a population with higher YLD rates at some ages can have the smaller MG under every life table if its lower rates at younger ages outweigh the excess. In Figure~\ref{fig:thm-common}, $A$'s YLD rates exceed $B$'s at ages 18--37 but its cumulative YLD rate is nowhere higher, so its MG, the average of the cumulative curve over the distribution of age at death, is no larger under any life table: under the two life tables whose densities of age at death are shown beneath panel (b), with life expectancies of 83.6 and 66.9 years, MG is 9.4 years for $A$ and 10.6 years for $B$ under the first and 6.9 and 7.7 years under the second.

Theorem~\ref{thm:common-mort} has the same form as dominance results for rank-dependent measures of health achievement, such as the achievement index of \citet{wagstaff2002}, with age in place of income rank, the cumulative YLD rate in place of the generalised concentration curve and survival in place of the weights that express aversion to inequality. The main practical use of the Theorem is that it tell us that comparing two cumulative YLD rates, as in Figure~\ref{fig:thm-common}(b), settles the comparison of MG under every life table at once without the need to compute MG under an infinite number of particular life tables.

\subsection{Comparing MG when only mortality differs}\label{sec:pure-age}

Comparing MG across mortality schedules requires an assumption about how YLD rates respond when mortality changes, because the age-specific YLD rates observed under one life table need not be those that would prevail under another. Here and in Section~\ref{sec:attrib} we make the assumption, usual in cause-deleted life tables \citep{nusselder1996}, that age-specific YLD rates are unaffected, and set aside the alternative that part of health loss is tied to the years before death \citep{riffe2017}, which microsimulation models that follow individuals through their life courses could accommodate \citep{guo2025,normatov2026}. The results are stated with weak inequalities, but since GBD's YLD rates are positive at every age, the inequalities of Theorem~\ref{thm:mortality} and Corollary~\ref{cor:attrib} hold strictly in applications whenever the survival curves being compared differ over a range of ages.

\begin{restatable}[Common YLD profile]{theorem}{thmmortality}\label{thm:mortality}
For two life tables $S^A$ and $S^B$:
\begin{enumerate}
\item[(a)] the following are equivalent:
\begin{enumerate}
\item[(i)] $G(S^A,y)\ge G(S^B,y)$ for every YLD profile $y$;
\item[(ii)] $S^A(a)\ge S^B(a)$ for every age $a$;
\end{enumerate}
\item[(b)] the following are equivalent:
\begin{enumerate}
\item[(i)] $G(S^A,y)\ge G(S^B,y)$ for every nondecreasing YLD profile $y$;
\item[(ii)] $T^A(a)\ge T^B(a)$ for every age $a$;
\end{enumerate}
\item[(c)] the following are equivalent:
\begin{enumerate}
\item[(i)] $\gb(S^A,y)\ge\gb(S^B,y)$ for every nondecreasing YLD profile $y$;
\item[(ii)] $T^A(a)/e_0^A\ge T^B(a)/e_0^B$ for every age $a$;
\end{enumerate}
\item[(d)] if $\mu^A(a)\le\mu^B(a)$ for every age $a$, conditions (a)(ii), (b)(ii) and (c)(ii) all hold, and so $G(S^A,y)\ge G(S^B,y)$ for every YLD profile $y$ and $\gb(S^A,y)\ge\gb(S^B,y)$ for every nondecreasing $y$.
\end{enumerate}
\end{restatable}

Theorem~\ref{thm:mortality} is an approximate counterpart of Theorem~\ref{thm:common-mort}, with the roles of the life table and the YLD profile exchanged. In the terms of the stochastic orders defined in Appendix~\ref{app:proofs}, conditions (a)(ii) and (b)(ii) say that the age at death under $S^A$ is larger than under $S^B$ in the usual stochastic order and in the increasing convex order respectively, the hypothesis of (d) says that it is larger in the hazard rate order, and condition (c)(ii) says that the life-table population of $A$ is larger than that of $B$ in the usual stochastic order. Part~(d) also means that an intervention that raises survival can widen MG even when it lowers YLD rates, so that MG and HALE can rank interventions in opposite orders.

\subsection{Contributions of risk factors to MG}\label{sec:attrib}

For any two populations, the decomposition of \citet{kitagawa1955} (Lemma~\ref{lem:kitagawa} in Appendix~\ref{app:proofs}) splits the difference in MG into a part due to YLD rates and a part due to mortality,
\begin{equation}\label{eq:decomp}
G^B-G^A=\int_0^\omega S^A(y^B-y^A)+\int_0^\omega(S^B-S^A)\,y^B .
\end{equation}

Let $O$, with life table $S^O$ and YLD profile $y^O$, be the observed population, and let $R$, with life table $S^R$ and YLD profile $y^R$, be the population with a risk factor removed, so that $y^O-y^R$ is the YLD rate attributable to the risk factor at each age. We assume that $R$ has the YLD profile $y^R$ under its own life table, so that $G^R=G(S^R,y^R)$.

\begin{definition}[Deletion effect and its components]\label{def:attrib}
The \emph{deletion effect} of the risk factor is $G^O-G^R$, the change in MG when the risk factor is removed. Its \emph{morbidity component} and \emph{mortality component} are
\begin{equation}\label{eq:Gamma}
\Gamma_{\mathrm{mb}}=\int_0^\omega S^O(y^O-y^R)\quad\text{and}\quad\Gamma_{\mathrm{mt}}=\int_0^\omega(S^O-S^R)\,y^R,
\end{equation}
so that, by \eqref{eq:decomp} with $A=O$ and $B=R$,
\begin{equation}\label{eq:delcomp}
G^O-G^R=\Gamma_{\mathrm{mb}}+\Gamma_{\mathrm{mt}}.
\end{equation}
\end{definition}

The morbidity component is the health loss that the attributable YLD rates account for under the observed life table, and the mortality component is the change in health loss, at the YLD rates $y^R$, that comes from the change in survival. GBD's contribution of a risk factor to MG, which it computes by multiplying life-table person-years in each age group by the YLD rates attributable to the risk factor, is the morbidity component alone. \citet{hay2026} present these contributions as shares of the observed MG and note that the decompositions are proportional and do not imply causality (p.~e503), so that GBD's contribution says how much of the observed MG the risk factor accounts for, while the deletion effect says how much MG would change if the risk factor were removed.

\begin{corollary}[Deletion effect and morbidity component]\label{cor:attrib}
If $S^R(a)\ge S^O(a)$ at every age, which holds in particular when $\mu^R(a)\le\mu^O(a)$ at every age, then $\Gamma_{\mathrm{mt}}\le0$, so that $G^O-G^R\le\Gamma_{\mathrm{mb}}$, with equality if and only if $y^R(a)=0$ at the ages where $S^R(a)>S^O(a)$.
\end{corollary}

\begin{figure}[!htbp]
\centering
\includegraphics[width=\textwidth]{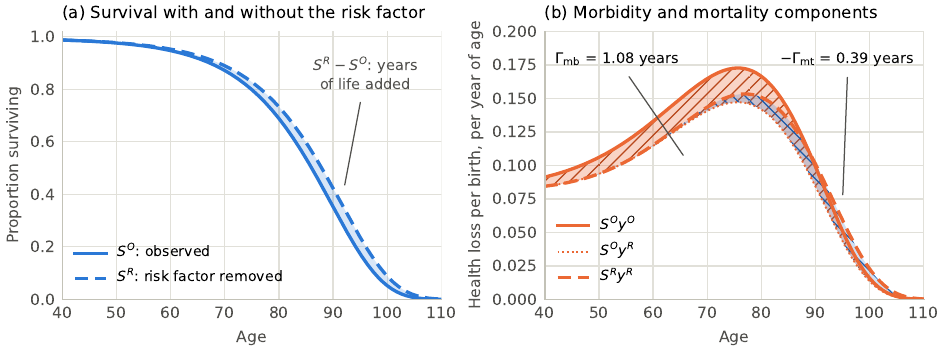}
\caption{Removing a hypothetical risk factor. (a) Survival $S^O$ with and $S^R$ without the risk factor. (b) Health loss per birth at each age, $S^Oy^O$, $S^Oy^R$ and $S^Ry^R$; the orange area is the morbidity component $\Gamma_{\mathrm{mb}}$ and the blue area is $-\Gamma_{\mathrm{mt}}$.}\label{fig:attrib}
\end{figure}

Corollary~\ref{cor:attrib} follows from \eqref{eq:delcomp}, since the integrand of $\Gamma_{\mathrm{mt}}$ is nonpositive when $S^R\ge S^O$ and is zero exactly where $S^R=S^O$ or $y^R=0$. Removing a risk factor that raises mortality therefore narrows MG by less than GBD's contribution, because the years of life that its removal adds are lived with some health loss. In Figure~\ref{fig:attrib} the risk factor accounts for up to 18\% of deaths and 15\% of YLDs at older ages, and removing it raises life expectancy from 83.6 to 85.3 years; $\Gamma_{\mathrm{mb}}$ is 1.08 years and $\Gamma_{\mathrm{mt}}$ is $-0.39$ years, so that the deletion effect is 0.69 years. The mortality component is small for risk factors that act mainly on disability and can be large for those that act mainly on mortality \citep{nusselder1996}, and for risk factors such as high body-mass index, which raise both, estimating it requires a risk-deleted life table. Applied to GBD's global estimates for 2023, with risk-deleted life tables constructed as in Section~\ref{sec:bmi} and YLD rates held fixed, the mortality components of tobacco, air pollution, high systolic blood pressure and dietary risks outweigh their morbidity components, so that removing any of them would widen MG, by between 0.10 and 0.48 years, although each is among the ten leading contributors to MG \citep{hay2026}; a ranking of risk factors by GBD's contribution can therefore differ from a ranking by deletion effect.

\section{The relation between MG and CY}\label{sec:cross}

MG is used as a policy indicator, and narrowing it is sometimes expected to reduce the cost of care \citep{saito2013}. The cost of care in a given year depends, however, on the morbidity of the people alive in that year, which CY measures. Since MG weights YLD rates by the life table and CY weights them by the actual population, the two can move apart: as a population ages with its life table and YLD rates unchanged, MG stays the same while the CYR rises whenever YLD rates rise with age, and if YLD rates also fall a little, MG falls while CY rises. This section gives conditions under which the CYR lies above or below the LYR, and Section~\ref{sec:sg-crude} applies them to Singapore.

\subsection{Older populations}\label{sec:ageing}

\begin{definition}[Older and younger than the life-table population]\label{def:otilt}
A population is \emph{older than its life-table population} (OTILT) if $1-C(a)\ge T(a)/e_0$ at every age, that is, if for every age $a$ the share of the population older than $a$ is at least the share of its life-table population older than $a$, and \emph{younger than its life-table population} (YTILT) if $1-C(a)\le T(a)/e_0$ at every age.
\end{definition}

An OTILT population has a mean age at least that of its life-table population, by Lemma~\ref{lem:icx}(a) and~(b) (though a population with a higher mean age need not be OTILT).

\begin{restatable}[CYR and LYR]{theorem}{thmwithin}\label{thm:within}
For any population:
\begin{enumerate}
\item[(a)] $\yb\ge\gb$ for every nondecreasing YLD profile if and only if the population is OTILT;
\item[(b)] $\yb\le\gb$ for every nondecreasing YLD profile if and only if the population is YTILT.
\end{enumerate}
\end{restatable}

\begin{figure}[!htbp]
\centering
\includegraphics[width=\textwidth]{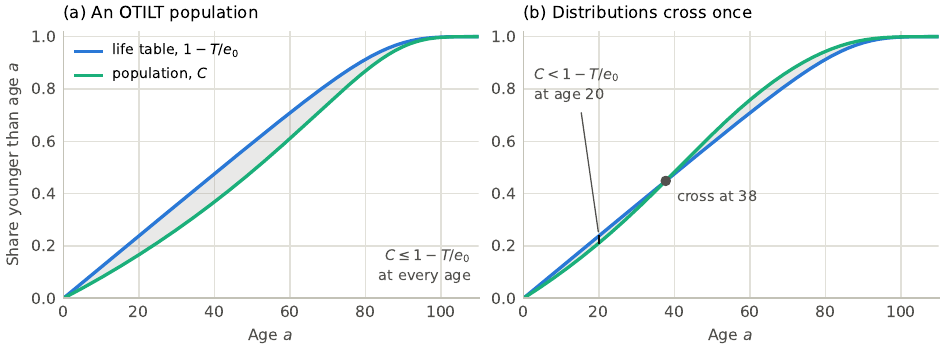}
\caption{Distribution functions $C$ of the population and $1-T/e_0$ of its life-table population \eqref{eq:ltpop}. (a) A population that is OTILT. (b) A population whose distribution function crosses that of its life-table population.}\label{fig:ageing}
\end{figure}

 In Figure~\ref{fig:ageing}(a) the population is OTILT. In panel (b) the population has relatively fewer children and fewer old people than its life-table population, as after a fertility transition; the distribution functions cross at about age 38.

Empirically, few populations are OTILT, mainly because the oldest people alive lived through higher mortality than the current life table assumes. In GBD 2023 estimates \citep{gbd2023results}, Japan is older than its life-table population at every age up to 85 but has about 0.2 percentage points fewer people over 90 and over 95, and Italy and Germany are older at every age up to 70 and 65 respectively.  On the other hand, Brazil, India and Nigeria are YTILT. 

\subsection{Convex YLD profiles}\label{sec:convex}

YLD profiles rise with age and are convex over adult ages, and for convex profiles a condition weaker than OTILT suffices. For any age $v$,
\begin{equation}\label{eq:excess}
\int_v^\omega\{1-C(a)\}\dd a=\int_{(v,\omega]}(a-v)\dd C(a)
\end{equation}
is the average number of years by which members of the population exceed age $v$, which is the share older than $v$ times their average excess over $v$; at $v=0$ it is the mean age.

\begin{definition}[More and fewer old years than the life table]\label{def:moytilt}
A population has \emph{more old years than its life table} (MOYTILT) if
\begin{equation}\label{eq:moytilt}
\int_v^\omega\{1-C(a)\}\dd a\;\ge\;\frac{1}{e_0}\int_v^\omega T(a)\dd a\quad\text{for every $v$,}
\end{equation}
and \emph{fewer old years than its life table} (FOYTILT) if
\begin{equation}\label{eq:foytilt}
\int_v^\omega\{1-C(a)\}\dd a\;\le\;\frac{1}{e_0}\int_v^\omega T(a)\dd a\quad\text{for every $v$.}
\end{equation}
\end{definition}
In the terms of \citet[chs.~1 and~4]{shaked2007}, OTILT and YTILT say that the age distribution of the population is larger or smaller than that of its life-table population in the usual stochastic order, and MOYTILT and FOYTILT that it is larger or smaller in the increasing convex order. An OTILT population is MOYTILT and a YTILT population is FOYTILT, but not conversely, since a population with fewer old people can have as many old years if those it has are older.

\begin{restatable}[Convex YLD profiles]{theorem}{thmconvex}\label{thm:convex}
For any population:
\begin{enumerate}
\item[(a)] $\yb\ge\gb$ for every nondecreasing convex YLD profile if and only if the population is MOYTILT;
\item[(b)] $\yb\le\gb$ for every nondecreasing convex YLD profile if and only if the population is FOYTILT.
\end{enumerate}
\end{restatable}

\begin{figure}[!htbp]
\centering
\includegraphics[width=\textwidth]{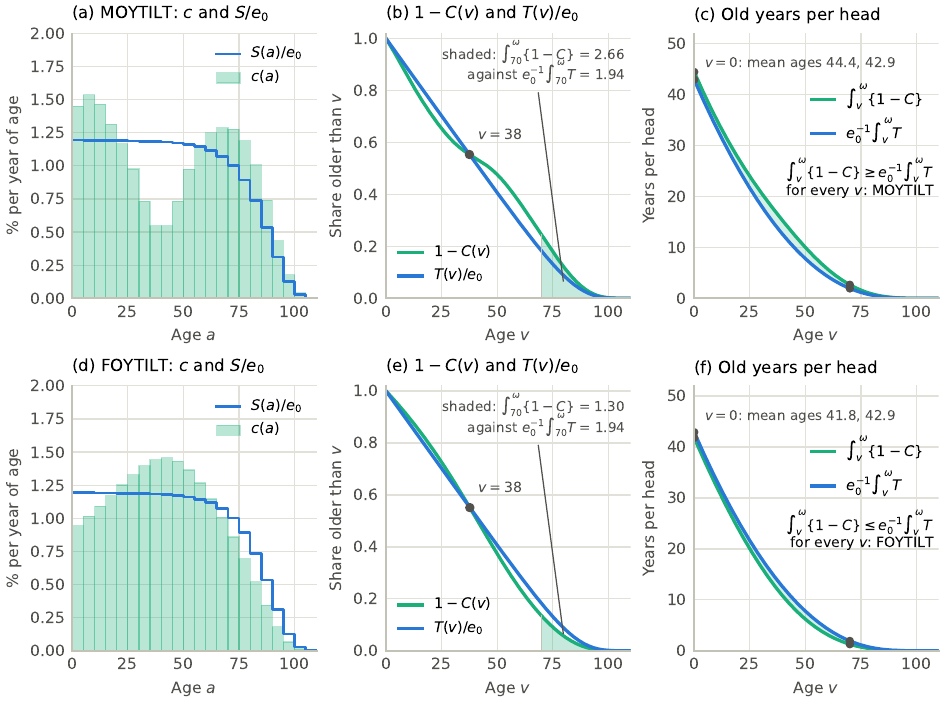}
\caption{Old years per head in two synthetic populations. Top row: a MOYTILT population. Bottom row: the FOYTILT population of Figure~\ref{fig:ageing}(b). (a), (d) Age densities $c$ of the population and $S/e_0$ of its life-table population, in five-year groups. (b), (e) Shares older than each age $v$, $1-C(v)$ and $T(v)/e_0$; the shaded area is $\int_{70}^\omega\{1-C\}$. (c), (f) The two sides of \eqref{eq:moytilt}, $\int_v^\omega\{1-C\}$ and $e_0^{-1}\int_v^\omega T$.}\label{fig:convex}
\end{figure}

The top row of Figure~\ref{fig:convex} shows a population with relatively more children and more old people than its life-table population (panel (a)). The shares of the two populations older than each age cross at about age 38 (panel (b)), so the population is not OTILT, but its old years per head are at least those of its life-table population whatever age is taken as the start of old age (panel (c)), so it is MOYTILT. Beyond age 70, for instance, it has 2.66 old years per head, since 24\% of it is older than 70, by 10.9 years on average, against 1.94 years for the life-table population, of which 18\% is older than 70, by 10.7 years on average. The bottom row shows the population of Figure~\ref{fig:ageing}(b), which has relatively fewer children and fewer old people than its life-table population (panel (d)). Its shares cross at the same age (panel (e)), so it is not YTILT, but its old years per head are at most those of its life-table population at every age (panel (f)), so it is FOYTILT; beyond age 70 it has 1.30 old years per head, with 13\% of it older than 70, by 9.7 years on average. At $v=0$ the curves in panels (c) and (f) are the mean ages.

For distributions where the population distribution and the life-table distribution exhibit the single-crossing property of Lemma~\ref{lem:cut}, FOYTILT and MOYTILT can be read off directly from a comparison of mean ages.  

\begin{restatable}[Single-crossing age structures]{corollary}{corsinglecross}\label{cor:singlecross}
For any population:
\begin{enumerate}
\item[(a)] Suppose that $1-C(a)-T(a)/e_0$ is nonpositive up to some age and nonnegative above it, so that the actual population has relatively more young and more old people than its life-table population. Then the population is MOYTILT if and only if its mean age is no lower than that of its life-table population, and in that case $\yb\ge\gb$ for every nondecreasing convex YLD profile.
\item[(b)] Suppose that $1-C(a)-T(a)/e_0$ is nonnegative up to some age and nonpositive above it, so that the actual population has relatively fewer young and fewer old people than its life-table population. Then the population is FOYTILT if and only if its mean age is no greater than that of its life-table population, and in that case $\yb\le\gb$ for every nondecreasing convex YLD profile.
\end{enumerate}
\end{restatable}

The crossing of part (b) of the corollary is the pattern that follows a fertility transition, with a bulge of middle-aged cohorts, and it is the pattern of Singapore (see discussion below).

\section{Illustration: Singapore}\label{sec:illustration}

In this section, we apply the results of the previous sections to interrogate the dynamics of the burden of disease in the Singapore population.  We use GBD 2023 estimates of deaths, YLDs and population by age \citep{gbd2023results}, which cover the total population, including non-citizens. We compute MG by Sullivan's method from an abridged life table built from GBD's death rates in 22 age groups, from under 1 to 95 and over, and from GBD's YLD rates in the same groups. Life expectancy rose from 75.6 years in 1990 to 85.4 in 2023 and MG from 8.90 to 11.25 years.   GBD's published MG estimates are slightly different, at 8.9 and 11.2 years. The small difference arises because GBD builds its life tables by single year of age, from mortality rates interpolated within age groups, which an abridged table built from the group rates cannot match exactly. We write $S^t$ and $y^t$ for the life table and YLD profile of year $t$, so that MG in year $t$ is $G(S^t,y^t)$. Throughout this section, the ages at which a difference between two cumulative curves changes sign are found by linear interpolation within age groups, as Remark~\ref{rem:abridged} allows. The data extraction, the analysis code and Figures~\ref{fig:thm-common} to~\ref{fig:singapore} were produced with the assistance of Claude (Anthropic), an AI system. The code, with a technical report that recomputes every computed number in the paper, is deposited at Zenodo \citep{morton2026code}, together with twelve checks by which a reader can confirm the GBD estimates it uses through IHME's GBD connector and the settings of the GBD Results tool queries that download them in full.

\subsection{The change in MG between 1990 and 2023}

\begin{figure}[!htbp]
\centering
\includegraphics[width=\textwidth]{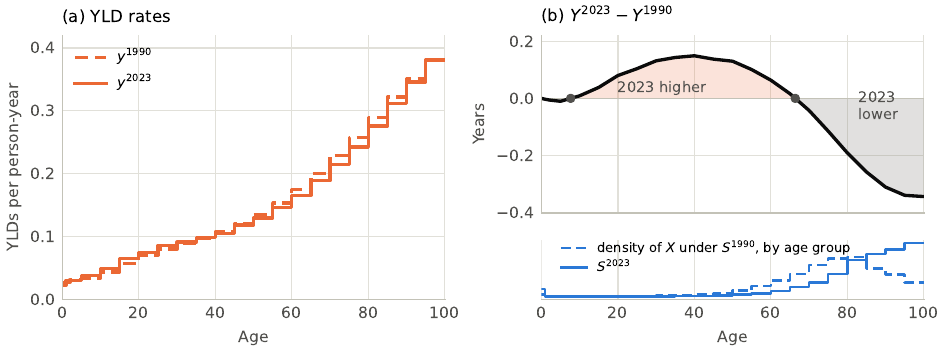}
\caption{Singapore, YLD rates in 1990 and 2023. (a) YLD profiles $y^{1990}$ and $y^{2023}$. (b) $Y^{2023}-Y^{1990}$, with the densities of age at death $X$ under $S^{1990}$ and $S^{2023}$, grouped by age, beneath; the open age group 95 and over is drawn over 95--100.}\label{fig:sg-rates}
\end{figure}

YLD rates changed little between 1990 and 2023 (Figure~\ref{fig:sg-rates}(a)): they were lower in 2023 below age 5, higher from 5 to 39 and lower from 40. Figure~\ref{fig:sg-rates}(b) plots $Y^{2023}-Y^{1990}$, the difference between the cumulative YLD rates of the two years, which is negative below age 7.7 and above age 66.4 and positive between.  Hence, by Theorem~\ref{thm:common-mort} neither year's YLD profile gives the smaller MG under every life table. By Lemma~\ref{lem:rep}, $G(S,y^{2023})-G(S,y^{1990})$ is the average of this curve over the distribution of age at death under $S$, shown beneath it for the two years' life tables. Since most deaths in both years occur after age 66, where $Y^{2023}$ is below $Y^{1990}$, the 2023 profile gives the smaller MG under both life tables: $G(S^t,y^{2023})-G(S^t,y^{1990})$ is $-0.14$ years for $t=1990$ and $-0.24$ years for $t=2023$. (Under a life table with enough deaths in middle age the ranking would be reversed.)  This hints that the increase in MG between 1990 and 2023 may be more attributable to changes in mortality patterns rather than a worsening of morbidity \textit{per se}.  

\begin{figure}[!htbp]
\centering
\includegraphics[width=\textwidth]{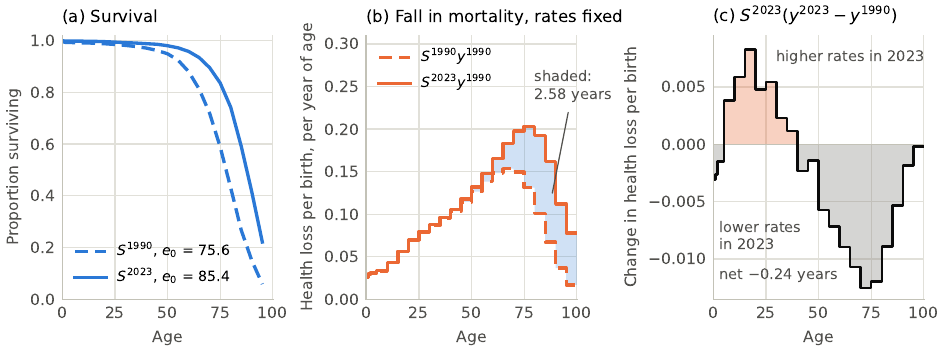}
\caption{Singapore, the change in MG between 1990 and 2023. (a) Survival $S^{1990}$ and $S^{2023}$. (b) Health loss per birth at each age at 1990 YLD rates, $S^{1990}y^{1990}$ and $S^{2023}y^{1990}$. (c) The change in YLD rates weighted by the 2023 life table, $S^{2023}(y^{2023}-y^{1990})$. The open age group 95 and over is drawn over 95--100.}\label{fig:sg-mort}
\end{figure}

Mortality fell at every age between 1990 and 2023 (Figure~\ref{fig:sg-mort}(a)) and the YLD rate rises with age in both years, so by Theorem~\ref{thm:mortality}(d) the fall in mortality alone widens MG and raises the LYR. Figure~\ref{fig:sg-mort}(b) shows where the widening comes from.  With YLD rates held at their 1990 values, the additional survivors add 2.58 years to MG, greater than the increase in the gap itself. Figure~\ref{fig:sg-mort}(c) shows the other term of the decomposition \eqref{eq:decomp}, $\int S^{2023}(y^{2023}-y^{1990})$, which adds a little health loss below age 40 and removes more above it, for a net change of in MG of $-0.24$ years. 

\subsection{Women and men}\label{sec:sex}

GBD reports its estimates by sex, as female or male, and we refer to the two groups as women and men, or as girls and boys where only children are meant. In GBD's estimates for 2023, women in Singapore had a life expectancy of 87.5 years and an MG of 12.1 years, and men 83.1 and 10.4 years. In our reconstruction the difference in MG is 1.73 years, which the decomposition \eqref{eq:decomp} splits into a part due to the difference in mortality and a part due to the difference in YLD rates. The split is not unique, since each of the two differences has to be weighted by one population or the other, so we report it both ways: weighting the difference in YLD rates by men's life table and the difference in survival by women's YLD rates gives 0.53 and 1.20 years, whereas the opposite choice of weights gives 0.50 and 1.23 years. Survival accounts for about 70\% of the difference under either weighting.

\begin{figure}[!htbp]
\centering
\includegraphics[width=\textwidth]{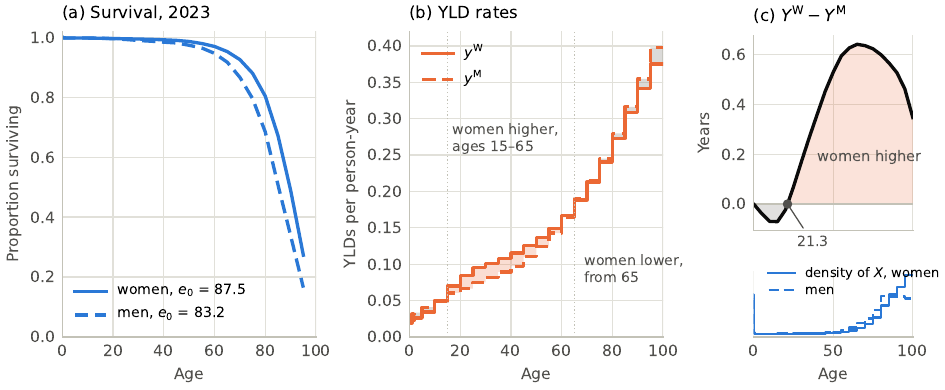}
\caption{Singapore, women and men in 2023. (a) Survival $S^{\mathrm{W}}$ and $S^{\mathrm{M}}$. (b) YLD profiles $y^{\mathrm{W}}$ and $y^{\mathrm{M}}$, shaded between them, in orange where women's rates are the higher and in grey where they are the lower. (c) $Y^{\mathrm{W}}-Y^{\mathrm{M}}$, which changes sign once, at age 21.3, with the densities of age at death $X$ under the two life tables beneath; the open age group 95 and over is drawn over 95--100.}\label{fig:sg-sex}
\end{figure}

Figure~\ref{fig:sg-sex}(a) shows the two life tables. Women's person-years lived above each age exceed men's at every age, and the share of their life-table population older than each age is likewise the larger.  Consequently, by Theorem~\ref{thm:mortality}(b) and (c) the female life table gives both a larger MG and a larger LYR than the male life table for every nondecreasing YLD profile. The stronger condition of Theorem~\ref{thm:mortality}(a) does not hold, because GBD gives girls slightly higher mortality than boys in the first month of life and at ages 2 to 4, so that women's survival is not above men's at every age. Women's YLD rates are below men's up to age 15 and from age 65 and above them between (Figure~\ref{fig:sg-sex}(b)), and their cumulative YLD rate is below men's up to about age 21.3 and above it from there to at least age 100 (Figure~\ref{fig:sg-sex}(c)), so by Lemma~\ref{lem:rep} women's YLD profile gives the larger MG under every life table in which all deaths occur between these ages, but by Theorem~\ref{thm:common-mort} not under every life table.

\subsection{High body-mass index}\label{sec:bmi}

\begin{figure}[!htbp]
\centering
\includegraphics[width=\textwidth]{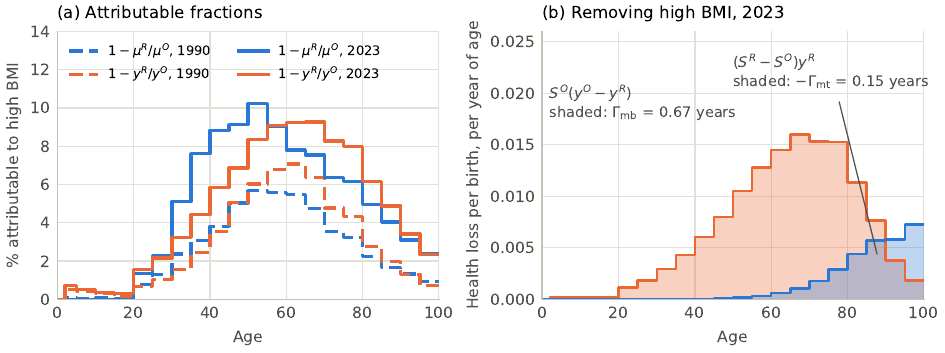}
\caption{Singapore, high body-mass index. (a) Percentages of deaths, $1-\mu^R/\mu^O$, and of YLDs, $1-y^R/y^O$, attributable to high body-mass index, in 1990 and 2023; GBD does not estimate these fractions below age 2, where we take them to be zero. (b) $S^O(y^O-y^R)$ and $(S^R-S^O)y^R$ in 2023, whose areas are $\Gamma_{\mathrm{mb}}$ and $-\Gamma_{\mathrm{mt}}$.}\label{fig:sg-bmi}
\end{figure}

By \eqref{eq:delcomp}, the deletion effect $G^O-G^R$ of a risk factor is the sum of a morbidity component $\Gamma_{\mathrm{mb}}$ and a mortality component $\Gamma_{\mathrm{mt}}$, and by Corollary~\ref{cor:attrib} the mortality component is at most zero when removing the risk factor lowers mortality at every age. Figure~\ref{fig:sg-bmi}(a) shows the fractions of deaths and YLDs that GBD attributes to high body-mass index. Both rose between 1990 and 2023 and in 2023 they were 5.62\% of all deaths and 6.35\% of all YLDs. 

We remove the risk factor by reducing deaths and YLDs at each age by these fractions, which compares the observed population with one whose exposure is at GBD's theoretical minimum-risk level. On this construction in 2023 life expectancy rises by 0.56 years and MG falls by 0.52 years. Figure~\ref{fig:sg-bmi}(b) splits the resulting change in MG: $\Gamma_{\mathrm{mb}}=0.67$ years and $\Gamma_{\mathrm{mt}}=-0.15$ years. The mortality component offsets 23\% of the morbidity component, against 30\% in 1990, when deaths made up a larger part of the burden of high body-mass index relative to YLDs. Appendix~\ref{app:bmi} sets out the construction in full and gives sensitivity bounds on both components. 

\subsection{MG and the CYR}\label{sec:sg-crude}

\begin{figure}[!htbp]
\centering
\includegraphics[width=\textwidth]{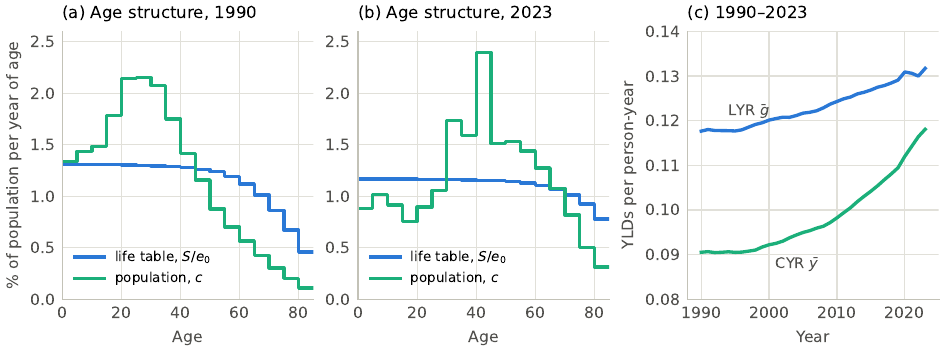}
\caption{Singapore, age structure and CY. (a), (b) Age densities $c$ of the population and $S/e_0$ of its life-table population in five-year groups, in 1990 and in 2023. The open age group 85 and over is left out, so that each closed group is drawn at its own density. (c) LYR $\gb$ and CYR $\yb$, 1990--2023.}\label{fig:singapore}
\end{figure}

Figure~\ref{fig:singapore}(a) and (b) compare the age structure of Singapore with that of its life-table population, in 1990 and in 2023. In 1990 the population held more people than its life-table population at every age below about 45 and fewer above, and by 2023 the children of 1990 had become the bulge of adults born before and during the fertility transition, so that Singapore now has relatively fewer children and fewer old people than its life-table population. In 1990 the share of the population younger than each age was at least that of its life-table population at every age, so that the population was YTILT and by Theorem~\ref{thm:within}(b) the CYR was at most the LYR for every nondecreasing YLD profile. By 2023 the difference $C(a)-1+T(a)/e_0$ is negative up to an age between 40 and 45 and positive above it, which is the crossing of Corollary~\ref{cor:singlecross}(b).  

The mean age, 41.9 years, is below the 43.8 years of the life-table population, so by the corollary the population is FOYTILT and $\yb\le\gb$ for every nondecreasing convex YLD profile. In fact, Singapore's YLD profile rises at every age and is close to convex above age 40 but not below, so the corollary does not by itself settle the comparison for the actual profile. Computing both quantities directly from GBD's age-specific YLD rates, life tables and population estimates gives a result consistent with the corollary: the CYR $\yb$ was 0.091 and the LYR $\gb$ 0.118 in 1990, and $\yb$ was 0.118 and $\gb$ 0.132 in 2023. Figure~\ref{fig:singapore}(c) shows the CYR below the LYR in every year between, with the difference roughly halved since 1990.

\section{Conclusion}\label{sec:conclusion}

MG addresses a question that the fertility transition and population ageing make more pressing.  Are the years of life that populations gain lived in good health, and can answers to this question be compared across populations whose age structures differ? Writing MG as the expected cumulative YLD rate at the life-table age at death (Lemma~\ref{lem:rep}) lets us use stochastic dominance results to interpret it. One YLD profile gives an MG and an LYR no larger than another's under every life table exactly when its cumulative YLD rate is nowhere higher (Theorem~\ref{thm:common-mort}), so comparisons of morbidity meant to be independent of mortality should rest on cumulative YLD rates. With age-specific YLD rates fixed, lower mortality at every age widens MG (Theorem~\ref{thm:mortality}), so that a widening MG is not by itself evidence of worsening non-fatal health. Since MG weights YLD rates by the life table and CY weights them by the actual population, MG can fall in an ageing population while CY (and the cost of care) rise. The CYR is at most the LYR for every rising YLD profile exactly when the population is YTILT, and at least the LYR exactly when it is OTILT, which few populations are at every age (Theorem~\ref{thm:within}).  For every rising convex profile the conditions are FOYTILT and MOYTILT (Theorem~\ref{thm:convex}), and after a fertility transition FOYTILT comes down to a comparison of mean ages (Corollary~\ref{cor:singlecross}(b)). In Singapore, with YLD rates fixed, the fall in mortality between 1990 and 2023 would on its own have widened MG by about 2.5 years.  This effect was mitigated because YLD rates fell at older ages.  

The conditions in these results can be checked from the rates, life tables and populations that GBD publishes.  Empirical studies could support interpretation of MG by reporting the curves on which the conditions rest, namely the cumulative YLD rate, the distribution of age at death and the age distribution of the population against that of its life-table population.  It would also be helpful to report, for risk factors that raise mortality, the mortality component of the deletion effect, computed on a risk-deleted life table, alongside GBD's contribution, as we have done for Singapore. Since Theorem~\ref{thm:common-mort}, which holds for every weighting of YLD rates that declines with age, is the counterpart of the dominance conditions that rank distributions of health for every achievement index in a family whose weights decline with rank \citep{wagstaff2002,makdissi2014}, methods developed for concentration indices could be carried over to MG. Since GBD's estimates are uncertain and comparisons of curves may fail narrowly or only over part of the age range, statistical tests for dominance \citep{davidson2000} and the machinery of almost stochastic dominance \citep{leshno2002} and restricted dominance \citep{atkinson1987} suggest how the results could be applied to estimated curves.

\section*{Data availability}

The GBD 2023 estimates used in this article are available from the GBD Results tool of the Institute for Health Metrics and Evaluation \citep{gbd2023results} under IHME's terms of use, which do not allow them to be redistributed, and so they are included neither with the article nor with its code. The code that reproduces every computed number and figure in the paper is deposited at Zenodo \citep{morton2026code}, \url{https://doi.org/10.5281/zenodo.24061704}, together with twelve checks by which a reader can confirm the estimates it uses through IHME's GBD connector and the query settings needed to download them in full; further information is available from the author on request. Source: Institute for Health Metrics and Evaluation. Used with permission. All rights reserved.

\section*{Acknowledgements}

The author thanks Simon Hay, Marie Ng and Borame Dickens for their helpful comments.

This work was funded by the National University of Singapore under the Academic Health Programme (AHP) through the POPUP project. AHP is a research programme administered and funded by National University of Singapore (NUS) that supports projects which align with shared strategic objectives between National University Health System Pte Ltd (NUHS) and NUS for education, research and clinical translation. The funder had no role in the study design, analysis, interpretation, writing or the decision to submit.

The concept and analysis plan are the author's idea. During the preparation of this work the author interacted intensively with Claude (Anthropic) to draft and edit text, develop and check mathematical proofs, and search for and check the literature. Claude also wrote the code that extracts the data, carries out the analysis and draws every figure in the paper, as described in Section~\ref{sec:illustration}. The author also used ChatGPT (OpenAI) to carry out an independent review of the paper, including the proofs and the code, which led to a number of small technical changes that did not materially affect the results or claims of the paper. The author reviewed and edited the content as needed and takes full responsibility for the content of the published article.

\appendix
\section{Proofs}\label{app:proofs}

\setcounter{equation}{0}
\renewcommand{\theequation}{A.\arabic{equation}}
\renewcommand{\theHequation}{A.\arabic{equation}}

Throughout, functions of age that differ only on a set of ages of total length zero have the same integrals and are treated as equal.

\paragraph{Stochastic orders.} Let $F$ and $H$ be distribution functions on $[0,\omega]$ with means $m_F$ and $m_H$. $F$ is larger than $H$ in the \emph{usual stochastic order} if $F\le H$ at every age, and in the \emph{increasing convex order} if
\begin{equation}\label{eq:icx}
\int_v^\omega\{1-F(a)\}\dd a\;\ge\;\int_v^\omega\{1-H(a)\}\dd a\quad\text{for every $v$.}
\end{equation}
If $F$ and $H$ have densities $\phi$ and $\psi$, $F$ is larger in the \emph{likelihood ratio order} if $\phi(v)\psi(u)\ge\phi(u)\psi(v)$ whenever $u<v$, which says that $\phi/\psi$ is nondecreasing where $\psi$ is positive. $F$ is larger than $H$ in the \emph{hazard rate order} if $\{1-F(v)\}\{1-H(u)\}\ge\{1-F(u)\}\{1-H(v)\}$ whenever $u<v$, which says that $(1-F)/(1-H)$ is nondecreasing where $1-H$ is positive, and, for distributions with mortality rates as in \eqref{eq:S}, that the mortality rate of $F$ is at most that of $H$ at every age. As in \eqref{eq:excess}, each side of \eqref{eq:icx} is the average number of years by which an age with that distribution exceeds $v$, and at $v=0$ it is the mean. For age at death in a life table, $1-F=S$, so that the left side of \eqref{eq:icx} is $T(v)$, and with equal means the increasing convex order is the convex order, in which the larger distribution is the more spread out \citep[chs.~1 and 4]{shaked2007}.

\begin{alemma}[Relations between stochastic orders]\label{lem:icx}
\begin{enumerate}
\item[(a)] If $F$ is larger than $H$ in the likelihood ratio order, it is larger in the hazard rate order; if it is larger in the hazard rate order, it is larger in the usual stochastic order; and if it is larger in the usual stochastic order, it is larger in the increasing convex order.
\item[(b)] If $F$ is larger than $H$ in the increasing convex order, then $m_F\ge m_H$.
\item[(c)] If $m_F$ and $m_H$ are positive and $F$ is larger than $H$ in the hazard rate order, then the distribution with density $(1-F)/m_F$ is larger than the distribution with density $(1-H)/m_H$ in the likelihood ratio order.
\end{enumerate}
\end{alemma}

\begin{alemma}[Kitagawa decomposition]\label{lem:kitagawa}
For weights $w^A$ and $w^B$ and rates $r^A$ and $r^B$ on $[0,\omega]$,
\begin{equation}\label{eq:kitagawa}
\int_0^\omega w^Br^B-\int_0^\omega w^Ar^A=\int_0^\omega w^A(r^B-r^A)+\int_0^\omega(w^B-w^A)\,r^B .
\end{equation}
\end{alemma}

\begin{alemma}[Characterisations of stochastic orders]\label{lem:char}
Write $\int\varphi\dd F$ for the average of a function $\varphi$ over the distribution $F$.
\begin{enumerate}
\item[(a)] $F$ is larger than $H$ in the usual stochastic order if and only if $\int\varphi\dd F\ge\int\varphi\dd H$ for every nondecreasing function $\varphi$ on $[0,\omega]$.
\item[(b)] $F$ is larger than $H$ in the increasing convex order if and only if $\int\varphi\dd F\ge\int\varphi\dd H$ for every nondecreasing convex function $\varphi$ on $[0,\omega]$.
\end{enumerate}
Both statements remain true when $\varphi$ is restricted to functions with values in $[0,1]$.
\end{alemma}

Lemma~\ref{lem:kitagawa} follows by adding and subtracting $\int w^Ar^B$. The first term is the effect of the difference in rates at the weights of $A$, and the second the effect of the difference in weights at the rates of $B$. Taking the terms in the other order gives $\int w^B(r^B-r^A)+\int(w^B-w^A)r^A$, and averaging the two gives the symmetric decomposition of \citet{kitagawa1955}, in which each difference is evaluated at the average of the other factor. With $w=S$ and $r=y$, \eqref{eq:G} turns \eqref{eq:kitagawa} into \eqref{eq:decomp}, and Section~\ref{sec:illustration} reports both orders.

Lemma~\ref{lem:icx}(a) and~(b) and Lemma~\ref{lem:char} are standard results on stochastic orders \citep[sections~1.A--1.C and~4.A]{shaked2007}, so we do not prove them. Lemma~\ref{lem:icx}(c) holds because the constants $m_F$ and $m_H$ cancel from the inequality that defines the likelihood ratio order for the two densities, which leaves the inequality that defines the hazard rate order. Lemma~\ref{lem:cut} extends the cut criterion of \citet[Lemma~b, p.~1253]{karlin1963}, which concerns distributions with equal means, and we prove it below. In Lemma~\ref{lem:char}, \citet{shaked2007} allow every nondecreasing, or nondecreasing convex, $\varphi$ for which the averages exist. On $[0,\omega]$ such a $\varphi$ is bounded, and the inequality is unchanged when a constant is added to $\varphi$ or $\varphi$ is multiplied by a positive number, so restricting $\varphi$ to values in $[0,1]$ loses nothing.

\begin{alemma}[Single crossing]\label{lem:cut}
If $F-H$ is nonnegative up to some age and nonpositive above it, then $F$ is larger than $H$ in the increasing convex order if and only if $m_F\ge m_H$.
\end{alemma}

\begin{proof}Write $D(v)$ for the left side of \eqref{eq:icx} minus the right side, so that $D(v)=\int_v^\omega(H-F)$, $F$ is larger than $H$ in the increasing convex order exactly when $D(v)\ge0$ for every $v$, and, since at $v=0$ each side of \eqref{eq:icx} is a mean, $D(0)=m_F-m_H$. Assume first that $F$ is larger than $H$ in the increasing convex order. Then $m_F\ge m_H$ by Lemma~\ref{lem:icx}(b). Conversely, assume $m_F\ge m_H$, and let $a^*$ be an age such that $F-H$ is nonnegative up to $a^*$ and nonpositive above it. For $v\ge a^*$, the integrand $H-F$ is nonnegative on $(v,\omega]$, so $D(v)\ge0$. For $v<a^*$, splitting the integral that defines $D(0)$ at $v$ gives $D(0)=\int_0^v(H-F)+\int_v^\omega(H-F)=\int_0^v(H-F)+D(v)$, so that $D(v)=D(0)+\int_0^v(F-H)$, and this last integral is nonnegative because $F-H\ge0$ on $[0,v]$, which gives $D(v)\ge D(0)=m_F-m_H\ge0$.
\end{proof}

\begin{lemmaone}[MG as an expectation over age at death]
For every life table $S$ and YLD profile $y$,
\begin{equation*}
G=\E[Y(X)]\tag{\ref{eq:rep}}
\end{equation*}
and, for every $a$ with $S(a)>0$,
\begin{equation*}
G_a=\E[Y(X)-Y(a)\mid X>a].\tag{\ref{eq:repa}}
\end{equation*}
\end{lemmaone}

\begin{proof}By \eqref{eq:Y}, $Y(X)=\int_0^\omega\mathbf{1}\{a<X\}\,y(a)\dd a$. Since $y\ge0$ (Definition~\ref{def:pop}), Tonelli's theorem allows the expectation and the integral to be exchanged, and \eqref{eq:X} then gives
\begin{equation}\label{eq:pf-rep}
\E[Y(X)]=\int_0^\omega\Pr(X>a)\,y(a)\dd a=\int_0^\omega S(a)y(a)\dd a ,
\end{equation}
which is $G$ by \eqref{eq:G}. For an age $a$ with $S(a)>0$, the same argument gives $\E[\{Y(X)-Y(a)\}\mathbf{1}\{X>a\}]=\int_a^\omega Sy$, and dividing by $S(a)=\Pr(X>a)$ gives \eqref{eq:repa} by \eqref{eq:Ga}.
\end{proof}

\thmcommonmort*

\begin{proof}(i)$\Leftrightarrow$(ii): For every life table $S$, \eqref{eq:gbar} gives $\gb(S,y^A)-\gb(S,y^B)=\{G(S,y^A)-G(S,y^B)\}/e_0$, and $e_0>0$ by \eqref{eq:e0}, since $S$ is right-continuous with $S(0)=1$ (Definition~\ref{def:lifetable}).

(iii)$\Rightarrow$(i): Assume (iii), so that $Y^A(a)\le Y^B(a)$ at every age, and let $S$ be a life table. By Lemma~\ref{lem:rep}, $G(S,y^A)-G(S,y^B)=\E[Y^A(X)-Y^B(X)]$, which is nonpositive because $Y^A(X)-Y^B(X)\le0$ whatever the value of $X$.

(i)$\Rightarrow$(iii): Assume (i), and let $a\in(0,\omega]$. The function $S=\mathbf{1}_{[0,a)}$ is a life table by Definition~\ref{def:lifetable}, one in which everyone dies at age $a$, and for it \eqref{eq:G} and \eqref{eq:Y} give $G(S,y)=\int_0^ay=Y(a)$ for every YLD profile $y$. Applying (i) to this life table gives $Y^A(a)\le Y^B(a)$, and at $a=0$ both sides are zero by \eqref{eq:Y}. If only life tables with continuous mortality rates are admitted, the same conclusion follows by approximating $\mathbf{1}_{[0,a)}$ by such life tables.

For the final statement, assume first that $y^A\le y^B$ at every age, and let $S$ be a life table and $a$ an age with $S(a)>0$. By \eqref{eq:Ga}, $G_a(S,y^A)-G_a(S,y^B)=S(a)^{-1}\int_a^\omega S\,(y^A-y^B)$, which is nonpositive because $S\ge0$. Conversely, assume that $G_a(S,y^A)\le G_a(S,y^B)$ for every life table $S$ and every $a$ with $S(a)>0$, and let $0\le a<b\le\omega$. The life table $S=\mathbf{1}_{[0,b)}$ has $S(a)=1$, and \eqref{eq:Ga} gives $G_a(S,y)=\int_a^by$, so that $\int_a^b(y^A-y^B)\le0$. Since this holds for all $a<b$, the Lebesgue differentiation theorem gives $y^A\le y^B$ at almost every age, and so, under the convention at the start of this appendix, at every age.
\end{proof}

\thmmortality*

\begin{proof}(a) Assume (ii), so that $S^A\ge S^B$ at every age. By \eqref{eq:G}, $G(S^A,y)-G(S^B,y)=\int(S^A-S^B)y$, which is nonnegative for every YLD profile because $y\ge0$ (Definition~\ref{def:pop}), so (i) holds. Conversely, assume (i), and suppose that (ii) fails, so that $S^A(a)<S^B(a)$ at some age $a$. Since $S^A$ and $S^B$ are right-continuous (Definition~\ref{def:lifetable}), $S^A<S^B$ on an interval $[a,a+\delta)$ with $\delta>0$, so the set $E$ of ages at which $S^A<S^B$ has positive length and $\int_E(S^B-S^A)>0$. The function $y=\mathbf{1}_E$ is a YLD profile (Definition~\ref{def:pop}), and for it \eqref{eq:G} gives $G(S^A,y)-G(S^B,y)=-\int_E(S^B-S^A)<0$, which contradicts (i). So (ii) holds.

(b) Write $X^A$ and $X^B$ for the ages at death under $S^A$ and $S^B$ \eqref{eq:X}. By \eqref{eq:X} and \eqref{eq:T}, $T(v)=\int_v^\omega\Pr(X>a)\dd a$, so by \eqref{eq:icx} condition (ii) says that the distribution of $X^A$ is larger than that of $X^B$ in the increasing convex order. Assume (ii), and let $y$ be a nondecreasing YLD profile. By \eqref{eq:Y}, $Y$ is then nondecreasing and convex, so Lemma~\ref{lem:char}(b) gives $\E[Y(X^A)]\ge\E[Y(X^B)]$, which by Lemma~\ref{lem:rep} is $G(S^A,y)\ge G(S^B,y)$, so (i) holds. Conversely, assume (i), and let $v\in[0,\omega)$. The function $y=\mathbf{1}_{[v,\omega]}$ is a nondecreasing YLD profile, and for it \eqref{eq:G} and \eqref{eq:T} give $G(S,y)=\int_v^\omega S=T(v)$, so (i) gives $T^A(v)\ge T^B(v)$, which is (ii).

(c) By \eqref{eq:gbar}, $\gb(S,y)$ is the average of $y$ over the life-table population, whose distribution function is $1-T/e_0$ by \eqref{eq:ltpop}. Condition (ii) therefore says that $1-T^A/e_0^A\le1-T^B/e_0^B$ at every age, that is, that the life-table population of $A$ is larger than that of $B$ in the usual stochastic order. The nondecreasing YLD profiles are the nondecreasing functions on $[0,\omega]$ with values in $[0,1]$ (Definition~\ref{def:pop}). Assume (ii). Then Lemma~\ref{lem:char}(a) gives $\gb(S^A,y)\ge\gb(S^B,y)$ for every nondecreasing YLD profile, which is (i). Conversely, assume (i). Then Lemma~\ref{lem:char}(a), with $\varphi$ restricted to values in $[0,1]$, gives that the life-table population of $A$ is larger than that of $B$ in the usual stochastic order, which is (ii).

(d) Assume $\mu^A\le\mu^B$ at every age. By \eqref{eq:S}, $S^A(a)/S^B(a)=\exp[\int_0^a\{\mu^B(v)-\mu^A(v)\}\dd v]$ wherever $S^B(a)>0$, which is nondecreasing in $a$ because the integrand is nonnegative. Hence $S^A(v)S^B(u)\ge S^A(u)S^B(v)$ whenever $u<v$, trivially when $S^B(v)=0$, so that, by \eqref{eq:X}, $X^A$ is larger than $X^B$ in the hazard rate order. By Lemma~\ref{lem:icx}(a), $X^A$ is then larger than $X^B$ in the usual stochastic order, which by \eqref{eq:X} is (a)(ii), and in the increasing convex order, which, as shown in the proof of (b), is (b)(ii). To compare the life-table populations, take $F$ and $H$ in Lemma~\ref{lem:icx}(c) to be the distribution functions of $X^A$ and $X^B$. Then $1-F=S^A$ and $1-H=S^B$ by \eqref{eq:X}, and $m_F=e_0^A$ and $m_H=e_0^B$ by \eqref{eq:e0}, which are positive as shown in the proof of Theorem~\ref{thm:common-mort}, so that the distributions with densities $(1-F)/m_F=S^A/e_0^A$ and $(1-H)/m_H=S^B/e_0^B$ are the life-table populations of $A$ and $B$ (Definition~\ref{def:lifetable}). Since $X^A$ is larger than $X^B$ in the hazard rate order, Lemma~\ref{lem:icx}(c) gives that the life-table population of $A$ is larger than that of $B$ in the likelihood ratio order, and so, by Lemma~\ref{lem:icx}(a), in the usual stochastic order, which, as shown in the proof of (c), is (c)(ii). The final statements follow from (a) and (c).
\end{proof}

\thmwithin*

\begin{proof}By \eqref{eq:yb} and \eqref{eq:gbar}, $\yb$ and $\gb$ are the averages of $y$ over the population, whose distribution function is $C$ (Definition~\ref{def:pop}), and over its life-table population, whose distribution function is $1-T/e_0$ by \eqref{eq:ltpop}. The nondecreasing YLD profiles are the nondecreasing functions on $[0,\omega]$ with values in $[0,1]$ (Definition~\ref{def:pop}).

(a) Assume the population is OTILT, so that $1-C\ge T/e_0$, that is, $C\le1-T/e_0$, at every age (Definition~\ref{def:otilt}). Then the population is larger than its life-table population in the usual stochastic order, and Lemma~\ref{lem:char}(a) gives $\yb\ge\gb$ for every nondecreasing YLD profile. Conversely, assume $\yb\ge\gb$ for every nondecreasing YLD profile. Then Lemma~\ref{lem:char}(a), with $\varphi$ restricted to values in $[0,1]$, gives that the population is larger than its life-table population in the usual stochastic order, so that $C\le1-T/e_0$ at every age, and the population is OTILT.

(b) The proof is the same as for (a), with the population and its life-table population exchanged, since being YTILT means that $C\ge1-T/e_0$ at every age (Definition~\ref{def:otilt}), that is, that the life-table population is larger than the population in the usual stochastic order.
\end{proof}

\thmconvex*

\begin{proof}As in the proof of Theorem~\ref{thm:within}, $\yb$ and $\gb$ are the averages of $y$ over the population and over its life-table population, with distribution functions $C$ and $1-T/e_0$, and the nondecreasing convex YLD profiles are the nondecreasing convex functions on $[0,\omega]$ with values in $[0,1]$. With $F=1-T/e_0$, the side of \eqref{eq:icx} for the life-table population is $\int_v^\omega(1-F)=e_0^{-1}\int_v^\omega T$, so the population is larger than its life-table population in the increasing convex order exactly when \eqref{eq:moytilt} holds, and smaller exactly when \eqref{eq:foytilt} holds.

(a) Assume the population is MOYTILT, so that \eqref{eq:moytilt} holds (Definition~\ref{def:moytilt}). Then the population is larger than its life-table population in the increasing convex order, and Lemma~\ref{lem:char}(b) gives $\yb\ge\gb$ for every nondecreasing convex YLD profile. Conversely, assume $\yb\ge\gb$ for every nondecreasing convex YLD profile. Then Lemma~\ref{lem:char}(b), with $\varphi$ restricted to values in $[0,1]$, gives that the population is larger than its life-table population in the increasing convex order, which is \eqref{eq:moytilt}, so the population is MOYTILT.

(b) The proof is the same as for (a), with the population and its life-table population exchanged, since being FOYTILT means that \eqref{eq:foytilt} holds (Definition~\ref{def:moytilt}), that is, that the life-table population is larger than the population in the increasing convex order.
\end{proof}

\corsinglecross*

\begin{proof}The distribution function of the population is $C$ (Definition~\ref{def:pop}) and that of its life-table population is $1-T/e_0$ by \eqref{eq:ltpop}. As shown in the proof of Theorem~\ref{thm:convex}, the population is larger than its life-table population in the increasing convex order exactly when \eqref{eq:moytilt} holds, that is, when it is MOYTILT, and smaller exactly when \eqref{eq:foytilt} holds, that is, when it is FOYTILT (Definition~\ref{def:moytilt}).

(a) Assume that $1-C-T/e_0$ is nonpositive up to an age $a^*$ and nonnegative above it. In Lemma~\ref{lem:cut}, take $F=C$ and $H=1-T/e_0$. Then $F-H=-(1-C-T/e_0)$ is nonnegative up to $a^*$ and nonpositive above it, so by Lemma~\ref{lem:cut} the population is larger than its life-table population in the increasing convex order, that is, it is MOYTILT, if and only if its mean age is at least that of its life-table population. Theorem~\ref{thm:convex}(a) then gives $\yb\ge\gb$ for every nondecreasing convex YLD profile.

(b) Assume that $1-C-T/e_0$ is nonnegative up to an age $a^*$ and nonpositive above it. In Lemma~\ref{lem:cut}, take $F=1-T/e_0$ and $H=C$. Then $F-H=1-T/e_0-C$ is nonnegative up to $a^*$ and nonpositive above it, so by Lemma~\ref{lem:cut} the life-table population is larger than the population in the increasing convex order, that is, the population is FOYTILT, if and only if the mean age of the life-table population is at least that of the population. Theorem~\ref{thm:convex}(b) then gives $\yb\le\gb$ for every nondecreasing convex YLD profile.
\end{proof}

\setcounter{equation}{0}
\renewcommand{\theequation}{\thesection.\arabic{equation}}
\renewcommand{\theHequation}{\thesection.\arabic{equation}}

\section{Construction of the figures}\label{app:figs}

Figures~\ref{fig:thm-common} to~\ref{fig:convex} use synthetic populations on ages $0$ to $\omega=110$. The integrals are computed by the trapezoidal rule on a grid of 0.05 years in Python, with code written with the assistance of Claude (Section~\ref{sec:illustration}) and deposited at Zenodo \citep{morton2026code}. Survival to age 110 is at most $10^{-3}$ in these life tables ($2.2\times10^{-4}$ in the base life table), and those who reach that age are treated as dying at it, so that $S(\omega)=0$ as Definition~\ref{def:lifetable} requires; the integrals of $S$ are unaffected. The densities of age at death drawn in Figure~\ref{fig:thm-common} are those of the continuous part of each distribution, scaled to a common height, and leave out the atom at age 110.

\paragraph{Base population.} The mortality rate is $\mu(a)=0.00015+8.5\times10^{-6}e^{0.105a}+0.003e^{-2a}$, which gives life expectancy 83.6 years and a standard deviation of age at death of 13.7 years. The YLD profile is $y(a)=0.08+0.24/\{1+e^{-(a-72)/11}\}$, which gives HALE 72.9 years and MG 10.63 years.

\paragraph{Figure~\ref{fig:thm-common}.} $y^B$ is the base profile and $y^A(a)=y^B(a)-0.04/\{1+e^{(a-15)/1.5}\}+0.04e^{-\{(a-27)/6\}^2}-0.03/\{1+e^{-(a-50)/6}\}$. The low-mortality life table is the base life table, and the high-mortality life table has $\mu(a)=0.002+5\times10^{-5}e^{0.09a}+0.05e^{-1.5a}+0.003e^{-\{(a-25)/8\}^2}$, with life expectancy 66.9 years.

\paragraph{Figure~\ref{fig:attrib}.} The observed population $O$ is the base population. The risk factor accounts for fractions $0.18/\{1+e^{-(a-45)/6}\}$ of deaths and $0.15/\{1+e^{-(a-40)/8}\}$ of YLDs at age $a$, and the population $R$ with the risk factor removed has $\mu^R$ and $y^R$ equal to $\mu^O$ and $y^O$ reduced by these fractions.

\paragraph{Figure~\ref{fig:ageing}.} Both populations have the base life table and age densities proportional to $S(a)w(a)$. In panel (a), $w(a)=e^{0.01a}$, which rises with age, so that the ratio of $c$ to the life-table age density $S/e_0$ is increasing and the population is OTILT by Lemma~\ref{lem:icx}(a). In panel (b), $w(a)=\exp\{-\int_0^a\rho(v)\dd v\}$ with $\rho(v)=-0.015+0.035/\{1+e^{-(v-45)/8}\}$, so that this ratio rises and then falls.

\paragraph{Figure~\ref{fig:convex}.} In the top row the population has the base life table and age density proportional to $S(a)w(a)$ with $w(a)=1-0.6e^{-\{(a-40)/13\}^2}+0.2e^{-\{(a-8)/8\}^2}+0.3/\{1+e^{-(a-70)/6}\}$, which gives relatively more children and more old people than the life-table population. The bottom row shows the population of panel (b) of Figure~\ref{fig:ageing}.

\section{Removing high body-mass index}\label{app:bmi}

Section~\ref{sec:bmi} removes high body-mass index by reducing deaths and YLDs at each age by the fractions that GBD attributes to it, which compares the observed population with one whose exposure is at GBD's theoretical minimum-risk level \citep[appendix~2, pp.~10--11]{gbd2023dirf}. Since the construction does not model an intervention that would bring this exposure about, and since it ignores competing risks, as cause-deleted life tables usually do, the risk-deleted life table should be read as a counterfactual reweighting rather than as a forecast. GBD does not estimate attributable fractions below age 2, where we take them to be zero.

Within the morbidity component $\Gamma_{\mathrm{mb}}=0.67$ years, most of the health loss falls at ages 45 to 85, whereas the years of life that removing the risk factor adds, which give $\Gamma_{\mathrm{mt}}=-0.15$ years, lie mostly above age 80. GBD's uncertainty intervals for the attributable fractions are wide, and recomputing with the lower and upper bounds of both fractions gives $\Gamma_{\mathrm{mb}}$ between 0.22 and 1.14 years, with a mortality component that offsets between 22\% and 36\% of it. These are sensitivity bounds rather than an uncertainty interval, since a proper interval would need the joint draws behind GBD's estimates \citep[250 draws;][appendix~2, p.~47]{gbd2023dirf}, which the GBD Results tool does not provide.

\bibliographystyle{rss-harv}
\bibliography{yld_gap}

\begin{thebibliography}{49}
\expandafter\ifx\csname natexlab\endcsname\relax\def\natexlab#1{#1}\fi
\providecommand{\url}[1]{\texttt{#1}}
\providecommand{\href}[2]{#2}
\providecommand{\path}[1]{#1}
\providecommand{\DOIprefix}{doi:}
\providecommand{\ArXivprefix}{arXiv:}
\providecommand{\URLprefix}{URL: }
\providecommand{\Pubmedprefix}{pmid:}
\providecommand{\doi}[1]{\href{http://dx.doi.org/#1}{\path{#1}}}
\providecommand{\Pubmed}[1]{\href{pmid:#1}{\path{#1}}}
\providecommand{\bibinfo}[2]{#2}
\ifx\xfnm\relax \def\xfnm[#1]{\unskip,\space#1}\fi
\bibitem[{Airoldi and Morton(2009)}]{airoldi2009}
\bibinfo{author}{Airoldi, M.} and \bibinfo{author}{Morton, A.}
  (\bibinfo{year}{2009}) \bibinfo{title}{Adjusting life for quality or
  disability: stylistic difference or substantial dispute?}
\newblock \bibinfo{journal}{Health Economics}, \bibinfo{volume}{18},
  \bibinfo{pages}{1237--1247}.
\newblock \DOIprefix\doi{10.1002/hec.1424}.
\bibitem[{Allison and Foster(2004)}]{allison2004}
\bibinfo{author}{Allison, R.A.} and \bibinfo{author}{Foster, J.E.}
  (\bibinfo{year}{2004}) \bibinfo{title}{Measuring health inequality using
  qualitative data}.
\newblock \bibinfo{journal}{Journal of Health Economics}, \bibinfo{volume}{23},
  \bibinfo{pages}{505--524}.
\newblock \DOIprefix\doi{10.1016/j.jhealeco.2003.10.006}.
\bibitem[{Anand and Hanson(1997)}]{anand1997}
\bibinfo{author}{Anand, S.} and \bibinfo{author}{Hanson, K.}
  (\bibinfo{year}{1997}) \bibinfo{title}{Disability-adjusted life years: a
  critical review}.
\newblock \bibinfo{journal}{Journal of Health Economics}, \bibinfo{volume}{16},
  \bibinfo{pages}{685--702}.
\newblock \DOIprefix\doi{10.1016/S0167-6296(97)00005-2}.
\bibitem[{Atkinson(1970)}]{atkinson1970}
\bibinfo{author}{Atkinson, A.B.} (\bibinfo{year}{1970}) \bibinfo{title}{On the
  measurement of inequality}.
\newblock \bibinfo{journal}{Journal of Economic Theory}, \bibinfo{volume}{2},
  \bibinfo{pages}{244--263}.
\newblock \DOIprefix\doi{10.1016/0022-0531(70)90039-6}.
\bibitem[{Atkinson(1987)}]{atkinson1987}
\bibinfo{author}{Atkinson, A.B.} (\bibinfo{year}{1987}) \bibinfo{title}{On the
  measurement of poverty}.
\newblock \bibinfo{journal}{Econometrica}, \bibinfo{volume}{55},
  \bibinfo{pages}{749--764}.
\newblock \DOIprefix\doi{10.2307/1911028}.
\bibitem[{Crimmins and Beltr{\'a}n-S{\'a}nchez(2011)}]{crimmins2011}
\bibinfo{author}{Crimmins, E.M.} and \bibinfo{author}{Beltr{\'a}n-S{\'a}nchez,
  H.} (\bibinfo{year}{2011}) \bibinfo{title}{Mortality and morbidity trends: is
  there compression of morbidity?}
\newblock \bibinfo{journal}{Journals of Gerontology Series B: Psychological
  Sciences and Social Sciences}, \bibinfo{volume}{66B},
  \bibinfo{pages}{75--86}.
\newblock \DOIprefix\doi{10.1093/geronb/gbq088}.
\bibitem[{Davidson and Duclos(2000)}]{davidson2000}
\bibinfo{author}{Davidson, R.} and \bibinfo{author}{Duclos, J.Y.}
  (\bibinfo{year}{2000}) \bibinfo{title}{Statistical inference for stochastic
  dominance and for the measurement of poverty and inequality}.
\newblock \bibinfo{journal}{Econometrica}, \bibinfo{volume}{68},
  \bibinfo{pages}{1435--1464}.
\newblock \DOIprefix\doi{10.1111/1468-0262.00167}.
\bibitem[{Foster and Shorrocks(1988)}]{foster1988}
\bibinfo{author}{Foster, J.E.} and \bibinfo{author}{Shorrocks, A.F.}
  (\bibinfo{year}{1988}) \bibinfo{title}{Poverty orderings}.
\newblock \bibinfo{journal}{Econometrica}, \bibinfo{volume}{56},
  \bibinfo{pages}{173--177}.
\newblock \DOIprefix\doi{10.2307/1911846}.
\bibitem[{Fries(1980)}]{fries1980}
\bibinfo{author}{Fries, J.F.} (\bibinfo{year}{1980}) \bibinfo{title}{Aging,
  natural death, and the compression of morbidity}.
\newblock \bibinfo{journal}{New England Journal of Medicine},
  \bibinfo{volume}{303}, \bibinfo{pages}{130--135}.
\newblock \DOIprefix\doi{10.1056/NEJM198007173030304}.
\bibitem[{Garmany and Terzic(2024)}]{garmany2024}
\bibinfo{author}{Garmany, A.} and \bibinfo{author}{Terzic, A.}
  (\bibinfo{year}{2024}) \bibinfo{title}{Global healthspan--lifespan gaps among
  183 {World Health Organization} member states}.
\newblock \bibinfo{journal}{JAMA Network Open}, \bibinfo{volume}{7},
  \bibinfo{pages}{e2450241}.
\newblock \DOIprefix\doi{10.1001/jamanetworkopen.2024.50241}.
\bibitem[{Garmany and Terzic(2025)}]{garmany2025}
\bibinfo{author}{Garmany, A.} and \bibinfo{author}{Terzic, A.}
  (\bibinfo{year}{2025}) \bibinfo{title}{Healthspan--lifespan gap differs in
  magnitude and disease contribution across world regions}.
\newblock \bibinfo{journal}{Communications Medicine}, \bibinfo{volume}{5},
  \bibinfo{pages}{381}.
\newblock \DOIprefix\doi{10.1038/s43856-025-01111-2}.
\bibitem[{{GBD 2023 Disease and Injury and Risk Factor
  Collaborators}(2025)}]{gbd2023dirf}
\bibinfo{author}{{GBD 2023 Disease and Injury and Risk Factor Collaborators}}
  (\bibinfo{year}{2025}) \bibinfo{title}{Burden of 375 diseases and injuries,
  risk-attributable burden of 88 risk factors, and healthy life expectancy in
  204 countries and territories, including 660 subnational locations,
  1990--2023: a systematic analysis for the {Global Burden of Disease Study}
  2023}.
\newblock \bibinfo{journal}{The Lancet}, \bibinfo{volume}{406},
  \bibinfo{pages}{1873--1922}.
\newblock \DOIprefix\doi{10.1016/S0140-6736(25)01637-X}.
\bibitem[{{Global Burden of Disease Collaborative
  Network}(2024)}]{gbd2023results}
\bibinfo{author}{{Global Burden of Disease Collaborative Network}}
  (\bibinfo{year}{2024}) \bibinfo{title}{{Global Burden of Disease Study 2023
  (GBD 2023) Results}}.
\newblock \bibinfo{howpublished}{Institute for Health Metrics and Evaluation,
  Seattle}.
\newblock \bibinfo{note}{\url{https://vizhub.healthdata.org/gbd-results/},
  accessed October 2026}.
\bibitem[{Gruenberg(1977)}]{gruenberg1977}
\bibinfo{author}{Gruenberg, E.M.} (\bibinfo{year}{1977}) \bibinfo{title}{The
  failures of success}.
\newblock \bibinfo{journal}{Milbank Memorial Fund Quarterly. Health and
  Society}, \bibinfo{volume}{55}, \bibinfo{pages}{3--24}.
\newblock \DOIprefix\doi{10.2307/3349592}.
\bibitem[{Guo et~al.(2025)Guo, Tysinger, Wee, Subramaniam, Ma, Ng
  et~al.}]{guo2025}
\bibinfo{author}{Guo, X.}, \bibinfo{author}{Tysinger, B.},
  \bibinfo{author}{Wee, H.L.}, \bibinfo{author}{Subramaniam, M.},
  \bibinfo{author}{Ma, S.}, \bibinfo{author}{Ng, T.P.} et~al.
  (\bibinfo{year}{2025}) \bibinfo{title}{Disease burden, lifetime healthcare
  cost and long-term intervention impact projections among older adults in
  {Singapore}}.
\newblock \bibinfo{journal}{Nature Aging}, \bibinfo{volume}{5},
  \bibinfo{pages}{1358--1369}.
\newblock \DOIprefix\doi{10.1038/s43587-025-00915-0}.
\bibitem[{Hadar and Russell(1969)}]{hadar1969}
\bibinfo{author}{Hadar, J.} and \bibinfo{author}{Russell, W.R.}
  (\bibinfo{year}{1969}) \bibinfo{title}{Rules for ordering uncertain
  prospects}.
\newblock \bibinfo{journal}{American Economic Review}, \bibinfo{volume}{59},
  \bibinfo{pages}{25--34}.
\bibitem[{Hay et~al.(2026)Hay, Nam, Saqib et~al.}]{hay2026}
\bibinfo{author}{Hay, S.I.}, \bibinfo{author}{Nam, P.}, \bibinfo{author}{Saqib,
  H.} et~al. (\bibinfo{year}{2026}) \bibinfo{title}{Global, regional, and
  national trends in the morbidity gap and contributing diseases, injuries, and
  risk factors, 1990--2023: a systematic analysis for the {Global Burden of
  Disease Study} 2023}.
\newblock \bibinfo{journal}{The Lancet Public Health}, \bibinfo{volume}{11},
  \bibinfo{pages}{e487--e505}.
\newblock \DOIprefix\doi{10.1016/S2468-2667(26)00098-8}.
\bibitem[{Hussain et~al.(2016)Hussain, J{\o}rgensen and
  {\O}sterdal}]{hussain2016}
\bibinfo{author}{Hussain, M.A.}, \bibinfo{author}{J{\o}rgensen, M.M.} and
  \bibinfo{author}{{\O}sterdal, L.P.} (\bibinfo{year}{2016})
  \bibinfo{title}{Refining population health comparisons: a multidimensional
  first order dominance approach}.
\newblock \bibinfo{journal}{Social Indicators Research}, \bibinfo{volume}{129},
  \bibinfo{pages}{739--759}.
\bibitem[{Issa et~al.(2024)Issa, Van~Ourti, van Baal and O'Donnell}]{issa2024}
\bibinfo{author}{Issa, J.}, \bibinfo{author}{Van~Ourti, T.},
  \bibinfo{author}{van Baal, P.} and \bibinfo{author}{O'Donnell, O.}
  (\bibinfo{year}{2024}) \bibinfo{title}{Ranking age-at-death distributions
  using dominance: robust evaluation of {United States} mortality trends,
  2006--2021}.
\newblock \bibinfo{journal}{Demography}, \bibinfo{volume}{61},
  \bibinfo{pages}{1143--1159}.
\newblock \DOIprefix\doi{10.1215/00703370-11460856}.
\bibitem[{Kakwani et~al.(1997)Kakwani, Wagstaff and van
  Doorslaer}]{kakwani1997}
\bibinfo{author}{Kakwani, N.}, \bibinfo{author}{Wagstaff, A.} and
  \bibinfo{author}{van Doorslaer, E.} (\bibinfo{year}{1997})
  \bibinfo{title}{Socioeconomic inequalities in health: measurement,
  computation, and statistical inference}.
\newblock \bibinfo{journal}{Journal of Econometrics}, \bibinfo{volume}{77},
  \bibinfo{pages}{87--103}.
\bibitem[{Karlin and Novikoff(1963)}]{karlin1963}
\bibinfo{author}{Karlin, S.} and \bibinfo{author}{Novikoff, A.}
  (\bibinfo{year}{1963}) \bibinfo{title}{Generalized convex inequalities}.
\newblock \bibinfo{journal}{Pacific Journal of Mathematics},
  \bibinfo{volume}{13}, \bibinfo{pages}{1251--1279}.
\newblock \DOIprefix\doi{10.2140/pjm.1963.13.1251}.
\bibitem[{Keyfitz and Caswell(2005)}]{keyfitz2005}
\bibinfo{author}{Keyfitz, N.} and \bibinfo{author}{Caswell, H.}
  (\bibinfo{year}{2005}) \bibinfo{title}{Applied Mathematical Demography}.
\newblock \bibinfo{edition}{Third} ed., \bibinfo{publisher}{Springer},
  \bibinfo{address}{New York}.
\bibitem[{Kitagawa(1955)}]{kitagawa1955}
\bibinfo{author}{Kitagawa, E.M.} (\bibinfo{year}{1955})
  \bibinfo{title}{Components of a difference between two rates}.
\newblock \bibinfo{journal}{Journal of the American Statistical Association},
  \bibinfo{volume}{50}, \bibinfo{pages}{1168--1194}.
\newblock \DOIprefix\doi{10.1080/01621459.1955.10501299}.
\bibitem[{Kramer(1980)}]{kramer1980}
\bibinfo{author}{Kramer, M.} (\bibinfo{year}{1980}) \bibinfo{title}{The rising
  pandemic of mental disorders and associated chronic diseases and
  disabilities}.
\newblock \bibinfo{journal}{Acta Psychiatrica Scandinavica},
  \bibinfo{volume}{62}, \bibinfo{pages}{382--397}.
\newblock \DOIprefix\doi{10.1111/j.1600-0447.1980.tb07714.x}.
\bibitem[{Leshno and Levy(2002)}]{leshno2002}
\bibinfo{author}{Leshno, M.} and \bibinfo{author}{Levy, H.}
  (\bibinfo{year}{2002}) \bibinfo{title}{Preferred by ``all'' and preferred by
  ``most'' decision makers: almost stochastic dominance}.
\newblock \bibinfo{journal}{Management Science}, \bibinfo{volume}{48},
  \bibinfo{pages}{1074--1085}.
\newblock \DOIprefix\doi{10.1287/mnsc.48.8.1074.169}.
\bibitem[{Makdissi and Yazbeck(2014)}]{makdissi2014}
\bibinfo{author}{Makdissi, P.} and \bibinfo{author}{Yazbeck, M.}
  (\bibinfo{year}{2014}) \bibinfo{title}{Measuring socioeconomic health
  inequalities in presence of multiple categorical information}.
\newblock \bibinfo{journal}{Journal of Health Economics}, \bibinfo{volume}{34},
  \bibinfo{pages}{84--95}.
\newblock \DOIprefix\doi{10.1016/j.jhealeco.2013.11.008}.
\bibitem[{Manton(1982)}]{manton1982}
\bibinfo{author}{Manton, K.G.} (\bibinfo{year}{1982}) \bibinfo{title}{Changing
  concepts of morbidity and mortality in the elderly population}.
\newblock \bibinfo{journal}{Milbank Memorial Fund Quarterly. Health and
  Society}, \bibinfo{volume}{60}, \bibinfo{pages}{183--244}.
\newblock \DOIprefix\doi{10.2307/3349767}.
\bibitem[{Mathers(1999)}]{mathers1999}
\bibinfo{author}{Mathers, C.D.} (\bibinfo{year}{1999}) \bibinfo{title}{Gains in
  health expectancy from the elimination of diseases among older people}.
\newblock \bibinfo{journal}{Disability and Rehabilitation},
  \bibinfo{volume}{21}, \bibinfo{pages}{211--221}.
\newblock \DOIprefix\doi{10.1080/096382899297639}.
\bibitem[{{Ministry of Health, Singapore}(2026)}]{moh2026}
\bibinfo{author}{{Ministry of Health, Singapore}} (\bibinfo{year}{2026})
  \bibinfo{title}{Gap between average and healthy life expectancy of
  {Singapore} residents and trends over past 10 years}.
\newblock \bibinfo{howpublished}{Reply to parliamentary question, 8 September
  2026}.
\newblock
  \bibinfo{note}{\url{https://www.moh.gov.sg/newsroom/gap-between-average-and-healthy-life-expectancy-of-singapore-residents-and-trends-over-past-10-years/},
  accessed 9 November 2026}.
\bibitem[{Mooney and Wiseman(2000)}]{mooney2000}
\bibinfo{author}{Mooney, G.} and \bibinfo{author}{Wiseman, V.}
  (\bibinfo{year}{2000}) \bibinfo{title}{Burden of disease and priority
  setting}.
\newblock \bibinfo{journal}{Health Economics}, \bibinfo{volume}{9},
  \bibinfo{pages}{369--372}.
\newblock
  \DOIprefix\doi{10.1002/1099-1050(200007)9:5<369::AID-HEC536>3.0.CO;2-D}.
\bibitem[{Moreno-Ternero et~al.(2023)Moreno-Ternero, Platz and
  {\O}sterdal}]{morenoternero2023}
\bibinfo{author}{Moreno-Ternero, J.D.}, \bibinfo{author}{Platz, T.T.} and
  \bibinfo{author}{{\O}sterdal, L.P.} (\bibinfo{year}{2023})
  \bibinfo{title}{{QALYs}, {DALYs}, and {HALYs}: a unifying framework for the
  evaluation of population health}.
\newblock \bibinfo{journal}{Journal of Health Economics}, \bibinfo{volume}{87},
  \bibinfo{pages}{102714}.
\bibitem[{Morton(2026)}]{morton2026code}
\bibinfo{author}{Morton, A.} (\bibinfo{year}{2026})
  \bibinfo{title}{{Interpreting the Morbidity Gap: Code and Technical Report}}.
\newblock \bibinfo{howpublished}{Zenodo}.
\newblock \DOIprefix\doi{10.5281/zenodo.24061704}.
\bibitem[{Murray and Acharya(1997)}]{murray1997}
\bibinfo{author}{Murray, C.J.L.} and \bibinfo{author}{Acharya, A.K.}
  (\bibinfo{year}{1997}) \bibinfo{title}{Understanding {DALYs}}.
\newblock \bibinfo{journal}{Journal of Health Economics}, \bibinfo{volume}{16},
  \bibinfo{pages}{703--730}.
\newblock \DOIprefix\doi{10.1016/S0167-6296(97)00004-0}.
\bibitem[{Murray et~al.(2012)Murray, Ezzati, Flaxman, Lim, Lozano, Michaud,
  Naghavi, Salomon, Shibuya, Vos, Wikler and Lopez}]{murray2012}
\bibinfo{author}{Murray, C.J.L.}, \bibinfo{author}{Ezzati, M.},
  \bibinfo{author}{Flaxman, A.D.}, \bibinfo{author}{Lim, S.},
  \bibinfo{author}{Lozano, R.}, \bibinfo{author}{Michaud, C.},
  \bibinfo{author}{Naghavi, M.}, \bibinfo{author}{Salomon, J.A.},
  \bibinfo{author}{Shibuya, K.}, \bibinfo{author}{Vos, T.},
  \bibinfo{author}{Wikler, D.} and \bibinfo{author}{Lopez, A.D.}
  (\bibinfo{year}{2012}) \bibinfo{title}{{GBD} 2010: design, definitions, and
  metrics}.
\newblock \bibinfo{journal}{The Lancet}, \bibinfo{volume}{380},
  \bibinfo{pages}{2063--2066}.
\bibitem[{Nord(2013)}]{nord2013}
\bibinfo{author}{Nord, E.} (\bibinfo{year}{2013}) \bibinfo{title}{Disability
  weights in the {Global Burden of Disease} 2010: unclear meaning and
  overstatement of international agreement}.
\newblock \bibinfo{journal}{Health Policy}, \bibinfo{volume}{111},
  \bibinfo{pages}{99--104}.
\newblock \DOIprefix\doi{10.1016/j.healthpol.2013.03.019}.
\bibitem[{Normatov et~al.(2026)Normatov, Song, Jin, Zhang, He, Lim, Zheng,
  Aziz, M{\"u}ller-Riemenschneider, Lee and Dickens}]{normatov2026}
\bibinfo{author}{Normatov, B.}, \bibinfo{author}{Song, H.},
  \bibinfo{author}{Jin, S.}, \bibinfo{author}{Zhang, X.}, \bibinfo{author}{He,
  Y.}, \bibinfo{author}{Lim, N.W.H.}, \bibinfo{author}{Zheng, V.F.},
  \bibinfo{author}{Aziz, M.H.B.M.},
  \bibinfo{author}{M{\"u}ller-Riemenschneider, F.}, \bibinfo{author}{Lee,
  C.P.L.} and \bibinfo{author}{Dickens, B.L.} (\bibinfo{year}{2026})
  \bibinfo{title}{Impact of the national steps challenge physical activity
  programme on obesity and type 2 diabetes prevalence in {Singapore} to 2050:
  simulation-based forecasting analysis}.
\newblock \bibinfo{journal}{The Lancet Regional Health -- Western Pacific},
  \bibinfo{volume}{68}, \bibinfo{pages}{101831}.
\newblock \DOIprefix\doi{10.1016/j.lanwpc.2026.101831}.
\bibitem[{{NUS Yong Loo Lin School of Medicine and Institute for Health Metrics
  and Evaluation}(2026)}]{nus2026}
\bibinfo{author}{{NUS Yong Loo Lin School of Medicine and Institute for Health
  Metrics and Evaluation}} (\bibinfo{year}{2026}) \bibinfo{title}{Press release
  on the {Global Burden of Disease} study of the morbidity gap in {Singapore}}.
\newblock \bibinfo{howpublished}{22 July 2026}.
\newblock
  \bibinfo{note}{\url{https://medicine.nus.edu.sg/wp-content/uploads/2026/07/2026-07-22-Press-release-NUS-IHME-GBD-Morbidity-in-Singapore_final.pdf},
  accessed 9 November 2026}.
\bibitem[{Nusselder et~al.(1996)Nusselder, van~der Velden, van Sonsbeek, Lenior
  and van~den Bos}]{nusselder1996}
\bibinfo{author}{Nusselder, W.J.}, \bibinfo{author}{van~der Velden, K.},
  \bibinfo{author}{van Sonsbeek, J.L.A.}, \bibinfo{author}{Lenior, M.E.} and
  \bibinfo{author}{van~den Bos, G.A.M.} (\bibinfo{year}{1996})
  \bibinfo{title}{The elimination of selected chronic diseases in a population:
  the compression and expansion of morbidity}.
\newblock \bibinfo{journal}{American Journal of Public Health},
  \bibinfo{volume}{86}, \bibinfo{pages}{187--194}.
\bibitem[{O'Donnell et~al.(2008)O'Donnell, van Doorslaer, Wagstaff and
  Lindelow}]{odonnell2008}
\bibinfo{author}{O'Donnell, O.}, \bibinfo{author}{van Doorslaer, E.},
  \bibinfo{author}{Wagstaff, A.} and \bibinfo{author}{Lindelow, M.}
  (\bibinfo{year}{2008}) \bibinfo{title}{Analyzing Health Equity Using
  Household Survey Data: A Guide to Techniques and Their Implementation}.
\newblock \bibinfo{publisher}{World Bank}, \bibinfo{address}{Washington, DC}.
\bibitem[{Preston et~al.(2001)Preston, Heuveline and Guillot}]{preston2001}
\bibinfo{author}{Preston, S.H.}, \bibinfo{author}{Heuveline, P.} and
  \bibinfo{author}{Guillot, M.} (\bibinfo{year}{2001})
  \bibinfo{title}{Demography: Measuring and Modeling Population Processes}.
\newblock \bibinfo{publisher}{Blackwell}, \bibinfo{address}{Oxford}.
\bibitem[{Riffe et~al.(2017)Riffe, van Raalte and Bijlsma}]{riffe2017}
\bibinfo{author}{Riffe, T.}, \bibinfo{author}{van Raalte, A.} and
  \bibinfo{author}{Bijlsma, M.J.} (\bibinfo{year}{2017})
  \bibinfo{title}{Healthy life expectancy, mortality, and age prevalence of
  morbidity}.
\newblock \bibinfo{type}{MPIDR Working Paper} \bibinfo{number}{WP-2017-015}.
  Max Planck Institute for Demographic Research.
\bibitem[{Rothschild and Stiglitz(1970)}]{rothschild1970}
\bibinfo{author}{Rothschild, M.} and \bibinfo{author}{Stiglitz, J.E.}
  (\bibinfo{year}{1970}) \bibinfo{title}{Increasing risk: {I}. {A} definition}.
\newblock \bibinfo{journal}{Journal of Economic Theory}, \bibinfo{volume}{2},
  \bibinfo{pages}{225--243}.
\newblock \DOIprefix\doi{10.1016/0022-0531(70)90038-4}.
\bibitem[{Saito(2013)}]{saito2013}
\bibinfo{author}{Saito, K.} (\bibinfo{year}{2013}) \bibinfo{title}{Aiming for a
  healthy {Japan}}.
\newblock \bibinfo{howpublished}{Keidanren (Japan Business Federation), May
  2013}.
\newblock
  \bibinfo{note}{\url{https://www.keidanren.or.jp/en/journal/2013/05_01kantougen.html},
  accessed 9 November 2026}.
\bibitem[{Salomon et~al.(2012)Salomon, Wang, Freeman, Vos, Flaxman, Lopez and
  Murray}]{salomon2012}
\bibinfo{author}{Salomon, J.A.}, \bibinfo{author}{Wang, H.},
  \bibinfo{author}{Freeman, M.K.}, \bibinfo{author}{Vos, T.},
  \bibinfo{author}{Flaxman, A.D.}, \bibinfo{author}{Lopez, A.D.} and
  \bibinfo{author}{Murray, C.J.L.} (\bibinfo{year}{2012})
  \bibinfo{title}{Healthy life expectancy for 187 countries, 1990--2010: a
  systematic analysis for the {Global Burden Disease Study} 2010}.
\newblock \bibinfo{journal}{The Lancet}, \bibinfo{volume}{380},
  \bibinfo{pages}{2144--2162}.
\bibitem[{Shaked and Shanthikumar(2007)}]{shaked2007}
\bibinfo{author}{Shaked, M.} and \bibinfo{author}{Shanthikumar, J.G.}
  (\bibinfo{year}{2007}) \bibinfo{title}{Stochastic Orders}.
\newblock \bibinfo{publisher}{Springer}, \bibinfo{address}{New York}.
\bibitem[{Wagstaff(2002)}]{wagstaff2002}
\bibinfo{author}{Wagstaff, A.} (\bibinfo{year}{2002})
  \bibinfo{title}{Inequality aversion, health inequalities and health
  achievement}.
\newblock \bibinfo{journal}{Journal of Health Economics}, \bibinfo{volume}{21},
  \bibinfo{pages}{627--641}.
\newblock \DOIprefix\doi{10.1016/S0167-6296(02)00006-1}.
\bibitem[{Wagstaff et~al.(1991)Wagstaff, Paci and van Doorslaer}]{wagstaff1991}
\bibinfo{author}{Wagstaff, A.}, \bibinfo{author}{Paci, P.} and
  \bibinfo{author}{van Doorslaer, E.} (\bibinfo{year}{1991}) \bibinfo{title}{On
  the measurement of inequalities in health}.
\newblock \bibinfo{journal}{Social Science \& Medicine}, \bibinfo{volume}{33},
  \bibinfo{pages}{545--557}.
\newblock \DOIprefix\doi{10.1016/0277-9536(91)90212-U}.
\bibitem[{Williams(1999)}]{williams1999}
\bibinfo{author}{Williams, A.} (\bibinfo{year}{1999})
  \bibinfo{title}{Calculating the global burden of disease: time for a
  strategic reappraisal?}
\newblock \bibinfo{journal}{Health Economics}, \bibinfo{volume}{8},
  \bibinfo{pages}{1--8}.
\newblock
  \DOIprefix\doi{10.1002/(SICI)1099-1050(199902)8:1<1::AID-HEC399>3.0.CO;2-B}.
\bibitem[{Wyper et~al.(2020)Wyper, Grant, Fletcher, McCartney, Fischbacher and
  Stockton}]{wyper2020}
\bibinfo{author}{Wyper, G.M.A.}, \bibinfo{author}{Grant, I.},
  \bibinfo{author}{Fletcher, E.}, \bibinfo{author}{McCartney, G.},
  \bibinfo{author}{Fischbacher, C.} and \bibinfo{author}{Stockton, D.L.}
  (\bibinfo{year}{2020}) \bibinfo{title}{How do world and {European} standard
  populations impact burden of disease studies? {A} case study of
  disability-adjusted life years ({DALYs}) in {Scotland}}.
\newblock \bibinfo{journal}{Archives of Public Health}, \bibinfo{volume}{78},
  \bibinfo{pages}{1}.

\end{thebibliography}

\end{document}